\documentclass[aps,preprint,nofootinbib,eqsecnum]{revtex4}
\usepackage{amsfonts}
\usepackage{amssymb}
\usepackage{amsmath}
\usepackage{graphicx}
\usepackage{amscd}

 \allowdisplaybreaks

\begin{document}

\title{String Theory in the Penrose Limit of AdS$_2 \times$S$^2$}
\author{\textbf{Cemsinan Deliduman}}
\affiliation{Feza G\"{u}rsey Institute, \c{C}engelk\"{o}y 34684, \.{I}stanbul, Turkey\\
\texttt{cemsinan@gursey.gov.tr}}
\author{\textbf{Burak T. Kaynak}}
\affiliation{Department of Physics, Bo\u{g}azi\c{c}i University, Bebek 34342, \.{I}%
stanbul, Turkey\\
\texttt{kaynakb@boun.edu.tr} \vspace{1.0cm}}

\begin{abstract}
The string theory in the Penrose limit of AdS$_{2}\times $S$^{2}$ is
investigated. The specific Penrose limit is the background known as the
Nappi-Witten spacetime, which is a plane-wave background with an axion
field. The string theory on it is given as the Wess-Zumino-Novikov-Witten
(WZNW) model on non-semi--simple group $H_{4}$. It is found that, in the
past literature, an important type of irreducible representations of the
corresponding algebra, $h_{4}$, were \emph{missed}. We present this \emph{%
\textquotedblleft new\textquotedblright } representations, which have the
type of \emph{continuous series representations}. All the three types of
representations of the previous literature can be obtained from the \emph{%
\textquotedblleft new\textquotedblright } representations by setting the
momenta in the theory to special values. Then we realized the affine
currents of the WZNW model in terms of four bosonic free fields and
constructed the spectrum of the theory by acting the negative frequency
modes of free fields on the ground level states in the $h_{4}$ continuous
series representation. The spectrum is shown to be free of ghosts, after the
Virasoro constraints are satisfied. In particular we argued that there is no
need for constraining one of the longitudinal momenta to have unitarity. The
tachyon vertex operator, that correspond to a particular state in the ground
level of the string spectrum, is constructed. The operator products of the
vertex operator with the currents and the energy-momentum tensor are shown
to have the correct forms, with the correct conformal weight of the vertex
operator.
\end{abstract}

\maketitle
\tableofcontents

\thispagestyle{empty}

\newpage

\section{Introduction}

The problem of how to define the string theory on a general curved manifold
and to determine the physical spectrum is a long standing problem, with only
partial successes. Definition of the string theory in a background
independent way is still elusive, and for specific backgrounds one has
different methods in one's disposal in order to determine the physical
spectrum and the associated physical quantities.

One specific kind of curved backgrounds are group manifolds. String theory
on a general group manifold is defined through the
Wess-Zumino-Novikov-Witten (WZNW) model \cite{Nov82},\cite{Wit84},\cite{GW86}%
. The power of this formulation comes from the fact that the string action
on such backgrounds has infinite dimensional current algebra as its symmetry
structure. The well--developed current algebra techniques allows one, in
principle, to find the physical spectrum of the theory, and then utilizing
the powerful conformal field theory (CFT) methods one can compute the
correlation functions and the other physically relevant quantities from the
theory. Straightforward as it might seem, however, things are not so when
one begins to analyze the WZNW model on various kinds of group manifolds.
The compact groups, e.g. SU(2) \cite{ZF86},\cite{GW86}, does not turn out to
be much problematic. Expanding the analytic symmetry currents in terms of
modes, one finds that these modes obey a Kac-Moody algebra \cite{GW86}. Then
one constructs the spectrum of the theory as infinite dimensional
representations of the Kac-Moody symmetry algebra of the theory. Since the
metric of the group manifold of a compact group is positive definite, one
does not confront with the problem of negative norm states. Seeing the
success of the method on SU(2) group manifold one expects the spectrum of
the WZNW model on a general compact group manifold can be determined with a
certain ease by using the know--how one gained from the analysis of WZNW
model on SU(2) group manifold.

When one wants to play the same game on a non-compact group manifold things
are not so easy however. Due to curved time coordinates, things get
complicated more than one initially assumes. The group manifold of the
non--compact simple group SL(2,R) is an example of this \cite{BRFW},\cite%
{B95},\cite{BDM}. Another example of a curved spacetime which is also a
manifold of a group is the four dimensional exact plane--wave background of
Nappi and Witten \cite{NW}. The group, whose manifold is the Nappi--Witten
(NW) spacetime, is a non-semi--simple group however. The corresponding
non-semi--simple algebra is the Heisenberg algebra with a rotation operator
added, which is denoted as $h_{4}$. It is also possible to view this algebra
as the centrally extended two dimensional Euclidean algebra, denoted as $%
E_{2}^{c}$. We denote the corresponding group as $H_{4}$.

The motivation to construct string theory on NW\ spacetime was just the one
of accomplishing the formulation of string theory on a manifold with curved
time and curved space coordinates. However, after the work of Berenstein,
Nastase and Maldacena (BMN) \cite{BMN} the motivation has changed. In \cite%
{BMN} it is conjectured that the string theory on a plane-wave background is
dual to some large charge limit of a gauge theory. This conjecture is
related to the AdS/CFT conjecture, since plane-wave backgrounds can be
obtained as Penrose limits \cite{penrose},\cite{guven},\cite{blau} of AdS
spacetimes. In the post--\cite{BMN} era the motivation became to determine
the string theory in NW spacetime and to check the BMN conjecture in this
most easy setting. It is possible to obtain the NW spacetime as the Penrose
limit of AdS$_{2}\times $S$^{2}$ spacetime as we will show in the next
section. It is conjectured in \cite{stro} that type 0A string theory on AdS$%
_{2}\times $S$^{2}$ is dual to a conformal quantum mechanics on the boundary
of AdS$_{2}$. In analogy with BMN\ limit, we expect the bosonic string
theory we quantize here would be dual to some large charge limit of
conformal quantum mechanics on circle.

Since Nappi and Witten's proposal of string theory on NW background as a
WZNW model on a non-semi--simple group \cite{NW} there have been a lot of
work on the solution of this $H_{4}$\ WZNW model \cite{KK}--\cite{KP}. The
main theme of all this works were the same. Find either quasi-free or free
field representations of the $\hat{h}_{4}$ current algebra and then built
the spectrum of string theory as an infinite dimensional representation of
the current algebra. In this approach, like in SU(2) case, the symmetry
currents are assumed to be analytical and expanded into modes. Then the
spectrum is built, as usual, by applying the negative frequency modes of the
currents on the ground level states, which constitute an irreducible
representation of the corresponding Lie algebra $h_{4}$. The irreducible
representations of the Lie algebra $h_{4}$ were determined in \cite{KK} and
the authors of that paper noted the extreme similarity of the irreducible
representations of $h_{4}$ with the irreducible representations of SL(2,R).
It turned out that $h_{4}$ has two infinite dimensional discrete series
representations which are conjugate to each other. Other than those, there
is a third, infinite dimensional continuous series representation. Since the
appearance of \cite{KK}, in all the subsequent works it is assumed that
these are the only possible irreducible representations of $h_{4}$ and there
are no other irreducible representations.

Specifically, from the point of view of dispersion relation, it seems that
in both discrete series representations the transverse momenta in the $x$
and $\bar{x}$ directions are not taken into account. Whereas in the specific
continuous series representation, since the momentum in $u$ direction is
taken equal to zero, there is only the contribution of transverse momenta,
both of which having the same value. Therefore, these representations
describe a string which moves in either of the two dimensional planes in NW
spacetime, either in $(x,\bar{x})$ plane or in $\left( u,v\right) $ plane.
Hence, none of these representations could adequately describe the true
motion of a string in NW background. In this paper we present an irreducible
representation of the Lie algebra $h_{4}$ which is \emph{missed} in \cite{KK}
and in all the subsequent publications on $H_{4}$ WZNW model. This
irreducible representation is the principal continuous series type, and
therefore it makes the similarity of irreducible representations of $h_{4}$
to those of SL(2,R) more pronounced. We also determined the conjugate
representation to the new continuous series representation. The specific
continuous series representation and both discrete series representations
presented in \cite{KK} are shown as special cases of the new representations
we found. The irreducible representations of the Lie algebra $h_{4}$ will be
presented in subsection \ref{IrReps}, together with the newly found ones.
Then we are going to construct the spectrum of the string following the
approach of Bars \cite{B95},\cite{BDM} on WZNW model on SL(2,R) manifold.
This spectrum is different than the previous proposals in the literature
\cite{KK}--\cite{KP}, but by setting appropriate momenta to zero, all the
other spectrums can be shown to be just special cases of the spectrum we
determined. We observed that there is no need to impose $p_{u}<1$ condition
claimed in \cite{macars},\cite{forgacs}--\cite{cheung}. We will comment on
the possible reason why this condition is encountered. After that we will
present the vertex operator, which corresponds to the states in the lowest
level of the infinite dimensional representation of the Kac--Moody algebra $%
\hat{h}_{4}$. We are going to make various checks in order to show that the
new representation of $h_{4}$ as the spectrum of quantum string in NW
spacetime is the correct choice. We will successfully show that the energy
of the ground level states and the conformal weight of the ground level
vertex operator are indeed equal to the dispersion relation, which is the
same as the negative of the eigenvalue of the second Casimir operator in the
\emph{new} irreducible representation.

Why were these continuous series type representations missed in the previous
analyses \cite{KK}--\cite{KP}? The reason could be that, when one writes the
action in terms of $\sigma -$model coordinates, one observes that the string
moving in NW\ spacetime feels a \textit{velocity dependent} harmonic
oscillator potential in the transverse coordinates \cite{HSPRD}. Due to this
\textit{velocity dependent} harmonic oscillator potential, one immediately
expects that the spectrum will be quantized, as in the particle case. Then,
it is further assumed that the spectrum also needs to contain a lowest
weight or highest weight state as in the particle case. However, this last
assumption is not correct in the string case. This is because the string
Hamiltonian is equal to $L_{0}+\bar{L}_{0}$, but \emph{not} to the number
operator. Therefore one does not need to stop at the state $\left\vert 0,%
\vec{p}\right\rangle $. States $\left\vert -n,\vec{p}\right\rangle $, with $%
n>0$, do not have negative energy, besides the same energy as the states $%
\left\vert n,\vec{p}\right\rangle $, in the string case.

How can we check independently that we found the correct spectrum? In the
case of string theory on flat spacetime, it is possible to impose the
light--cone gauge and analyze the solution of the string $\sigma -$model in
that gauge. It is shown by Horowitz and Steif in \cite{HSPRD} that in order
to be able to impose the string light--cone gauge on a curved background,
there should be a covariantly constant null vector on that curved
background. This means that the curved background should be a pp-wave
spacetime in order to implement the light--cone gauge of string theory.\ The
group manifolds of SU(2) and SL(2,R) do not contain such covariantly
constant null vector. Therefore, for those WZNW\ models we cannot make an
independent check of the spectrum through the string light--cone gauge.
Whereas in the case of $H_{4}$ WZNW model the string $\sigma -$model action
is solved in the light--cone gauge some time ago in \cite{forgacs}. However,
in these work as well, the authors needed to express the string spectrum as
a representation of the $\hat{h}_{4}$ Kac--Moody algebra, and they have used
the incomplete set of irreducible representations of the $h_{4}$ Lie algebra
similar to all the other works on $H_{4}$ WZNW model. This is one problem
with their result. The other one is that they could not write a modular
invariant one--loop partition function, even though it is expected that the
string light--cone will ensure the complete determination of the physical
spectrum of the quantum string, free of negative norm states. The problem
with the modular invariance is addressed, but not resolved in \cite{forgacs}%
. A possible reason for the appearance of such a problem is stated later in
\cite{HS}. These authors noticed that in the string light--cone gauge in NW
spacetime, the coordinates on the string worldsheet, $\sigma $ and $\tau $,
are treated non-equivalently. They claim that the modular invariance might
be attained if one treats them equally, possibly in covariant gauge.
Therefore, even though the string light--cone gauge is possible to implement
in the NW\ spacetime according to \cite{HSPRD}, the presence of the
background axion field gives rise to non-equivalent treatment of the string
worldsheet coordinates in the light--cone gauge and subsequently causes the
partition function not to be modular invariant. Thus it could be concluded
that even in the $H_{4}$ WZNW model, though implementable, the string
light--cone gauge has intrinsic problems and thus can not be assumed as the
guide to decide the correct physical spectrum of the quantum string.

Before presenting the details of our work, we outline and summarize the main
results. In the next section we will show that the NW spacetime is a Penrose
limit of AdS$_{2}\times $S$^{2}$ spacetime and then we are going to discuss
the WZNW\ model on it in general terms. We will discuss the irreducible
representations of the non-semi--simple algebra $h_{4}$ and present the
\textit{new} representations. The wave functional, which gives a functional
representation of the group corresponding to the \textit{new}
representations of the algebra, will be given in coherent state basis and
then the dispertion relation will be determined. In section III, we are
going to start with the free-field realization of the symmetry currents.
Then we will construct the spectrum by using the modes of free fields and
show that no ghosts remain in the spectrum after the Virasoro constraints
are satisfied. Depending on the values of transverse or longitudinal momenta
the string spectrum is described as one of the representations of Kac--Moody
algebra $\hat{h}_{4}$. After the determination of the string spectrum, we
will construct the \textquotedblleft tachyon\textquotedblright\ vertex
operator in position and momentum spaces in section IV. The form of vertex
operator in momentum space is useful to perform operator products with the
currents and the energy-momentum tensor. We will present the correct quantum
ordering of the vertex operator and show that it has correct conformal
dimension. We will conclude with a summary and comments on possible future
research.

\section{WZNW Model in the Nappi--Witten Spacetime}

\textit{Plane fronted waves with parallel rays} \cite{brinkmann},
abbreviated as \textit{pp-waves}, are the most general solutions to the
Einstein equations in four dimensions, with a covariantly constant null
Killing vector \cite{KSMH},\cite{HSPRD}. Their metric are generally given as%
\begin{equation}
ds^{2}=2dudv+dxd\bar{x}+f(u,x,\bar{x})du^{2}.  \label{pp-waveMetric}
\end{equation}%
If the function $f$ is quadratic in $x$ and $\bar{x}$, such spacetimes are
called exact plane waves \cite{baldwin},\cite{HSPRD}.

The Nappi--Witten (NW) spacetime is a special case of this, where the
function $f$ is also independent of $u$. It is the four dimensional exact
pp-wave background together with an axion field and can be obtained as a
Penrose limit of different geometries. It can be obtained as the Penrose
limit of the near horizon geometry of NS5 branes, $M_{6}\times R\times S^{3}$
\cite{KPR}, by booting along the null geodesic spinning along an equator of
the three sphere, or from the near horizon geometry of a NS5 brane wrapped
on $S^{2}$ \cite{GO}. Considering the NW spacetime as the Penrose limit of $%
M_{6}\times R\times S^{3}$, in \cite{KP} the string theory on NW spacetime
is also considered as the corresponding limit of $R\times SU(2)$ WZNW model
on $R\times S^{3}$. The\ level of WZNW model on $R$ is taken negative so as
to interpret this coordinate as the time coordinate. However, as will be
seen the zero-grade algebra, $h_{4}$, of the current algebra of WZNW\ model
on NW spacetime has some infinite dimensional representations which cannot
be thought as some contraction of representations of $U(1)\times SU(2)$
group. In this paper we are going to consider the NW spacetime as a Penrose
limit of AdS$_{2}\times $S$^{2}$ background. Since the quantization of
string theory on AdS$_{2}\times $S$^{2}$ is not known \cite{S98} we will not
be able to check our results by comparing them to some limit of the string
theory results on AdS$_{2}\times $S$^{2}$. However, the quantization of
string theory on NW spacetime may help to define string theory
perturbatively on AdS$_{2}\times $S$^{2}$.

\subsection{Geometry of the Penrose Limit of AdS$_{2}\times $S$^{2}$}

Depending on the choice of the null geodesic, the Penrose limit of AdS$%
_{2}\times $S$^{2}$ background turns out to be either Minkowski spacetime or
a plane--wave background \cite{blau}. The null geodesic, along which we are
going to boost the momentum of a test particle in order to obtain the NW
spacetime, passes through the center of AdS$_{2}$ and spins around the
equator of S$^{2}$. In global coordinates the metric of AdS$_{2}\times $S$%
^{2}$ is
\begin{equation}
ds^{2}=R\left( -\cosh ^{2}\rho \ dt^{2}+d\rho ^{2}+\cos ^{2}\theta \ d\phi
^{2}+d\theta ^{2}\right) .  \label{AdS2xS2}
\end{equation}%
where the coordinates $\left( \rho ,t\right) $ describe the AdS$_{2}$ part,
and the coordinates $\left( \theta ,\phi \right) $ describe the S$^{2}$ part
of the geometry. We take AdS$_{2}$ and S$^{2}$ parts to have the same radius
of curvature $R$. The Penrose limit is then the metric seen by a highly
boosted particle, that is it is moving in the vicinity of a null geodesic.
To find the specific Penrose limit we desire, we redefine the coordinates:
\begin{equation}
u=-\left( t+\phi \right) ,\quad v=\frac{R}{2}\left( t-\phi \right) ,\quad
x^{+}=\sqrt{R}\left( \rho +i\theta \right) ,\quad x^{-}=\sqrt{R}\left( \rho
-i\theta \right) .  \label{redef1}
\end{equation}%
and then take the $R\rightarrow \infty $ limit. Then the metric becomes
\begin{equation}
ds^{2}=2dudv+dx^{+}dx^{-}-x^{+}x^{-}du^{2}.  \label{nw4_1}
\end{equation}%
which has the standard form of metric of pp-wave backgrounds. To see that
this metric is the same as the metric of Nappi--Witten background as given
in \cite{NW} we do one more redefinitions:
\begin{equation}
x^{+}=e^{iu}\left( a_{1}+ia_{2}\right) \quad and\quad x^{-}=e^{-iu}\left(
a_{1}-ia_{2}\right) .  \label{redef2}
\end{equation}%
These turn the metric into
\begin{equation}
ds^{2}=2dudv+da_{k}da_{k}+\epsilon _{jk}da_{j}a_{k}du+bdu^{2},  \label{nw4_3}
\end{equation}%
which is the metric of the Nappi--Witten background \cite{NW}, after one
also performs the shift $v\rightarrow v+\frac{b}{2}u$. Nappi--Witten
background is not just a plane--wave background, there is also an
appropriate magnetic field (axion field), to cancel the energy of the wave
\cite{NW},\cite{KK}. Due to the presence of non-zero magnetic field in the
background, this spacetime sometimes is called as \textquotedblleft Hpp-wave
spacetime\textquotedblright\ \cite{bianchi}. Like metric, the antisymmetric $%
B_{\mu \nu }$ field may also be obtained via the same limiting procedure
from the NS field of an appropriate string theory model in AdS$_{2}\times $S$%
^{2}$ spacetime \cite{guven}. We will not concern with the form of that
string theory model in AdS$_{2}\times $S$^{2}$ spacetime in this paper. In
the following section we will describe the WZNW model in Nappi--Witten
background and read off the antisymmetric tensor field by identifying the
WZNW action with the $\sigma $--model action written in this background.

\subsection{$H_{4}$ WZNW Model}

String theory on group manifolds are analyzed through the WZNW model. The
action of the WZNW model consists of two parts. The first part is the
straightforward generalization of the Polyakov action of string theory on
flat spacetime into the curved background of the group manifold. This part
of the action is invariant under the corresponding Lie group. The second
part of the action, which is required in order to make the full action
invariant under the two dimensional conformal symmetry \cite{Wit84},
enhances the symmetry of the action from the Lie group to the corresponding
Kac--Moody group. Due to the equations of motion, the Noether currents that
one derives from the WZNW action separate into holomorphic and
anti-holomorphic parts. Each set of currents (holomorphic and
anti-holomorphic) separately obey an operator product algebra which is an
affine current algebra. Expanding the currents in terms of modes, one can
rewrite this current algebra as a Kac--Moody algebra. This Kac--Moody
algebra is nothing but the affinization of the Lie algebra associated to the
isometry group of that group manifold.

In this section we are going to review the WZNW model in the Nappi--Witten
background, which is first described in \cite{NW}. The $\sigma $--model in
Nappi--Witten background can be written as the WZNW model on the group
manifold of a non-semi--simple group. The corresponding non-semi--simple
algebra is the Heisenberg algebra with a rotation operator added, which
rotates both the momentum and the position operators to each other,
\begin{equation}
\left[ P_{1},P_{2}\right] =T,\quad \left[ J,P_{1}\right] =P_{2},\quad \left[
J,P_{2}\right] =-P_{1},\quad \left[ T,\ any\right] =0.  \label{h4_v1}
\end{equation}%
Under this view, the algebra is denoted as $h_{4}$, higher dimensional
generalizations, $h_{2n+2}$, of which contains $n$ momentum and $n$ position
operators. It is also possible to view this algebra as the centrally
extended two dimensional Euclidean algebra, denoted as $E_{2}^{c}$ \cite{KK}%
. The corresponding group can be denoted as either $H_{4}$ or NW group, to
emphasize that it is half of the isometry group of NW spacetime. In this
paper we will denote it as $H_{4}$ due to the future possibility of
extending this work to the case of higher Heisenberg groups $H_{2n+2}$ as in
\cite{bianchi}. For the manifold of non-semi--simple group $H_{4}$ we will
use the terms \textquotedblleft $H_{4}$ manifold\textquotedblright\ and
\textquotedblleft NW spacetime\textquotedblright\ randomly.

The WZNW action which describes the closed strings on $H_{4}$ manifold is
given by
\begin{equation}
S=\frac{k}{4\pi }\int_{\Sigma }d^{2}\sigma \ Tr\left( g^{-1}\partial
^{\alpha }g\ g^{-1}\partial _{\alpha }g\right) -\frac{k}{12\pi }%
\int_{B}d^{3}\varsigma \ \epsilon _{\alpha \beta \gamma }Tr\left(
g^{-1}\partial ^{\alpha }g\ g^{-1}\partial ^{\beta }g\ g^{-1}\partial
^{\gamma }g\right) .  \label{WZWaction}
\end{equation}%
where $g$ is the group element of $H_{4}$ and $g^{-1}dg=\left(
g^{-1}\partial _{\sigma }g\right) d\sigma +\left( g^{-1}\partial _{\tau
}g\right) d\tau $ is the pull-back of the left invariant one-form on the $%
H_{4}$ group manifold to the closed string worldsheet. To obtain the WZNW
action in terms of the coordinates of the background we parametrize the $%
H_{4}$ group element as in \cite{NW},
\begin{equation}
e^{a_{1}P^{1}+a_{2}P^{2}}e^{uJ+vT},
\end{equation}%
where all the coordinates, $a_{1},\ a_{2},\ u,\ v$, are real. In \cite{NW}
comparing the resulting form of the action with the $\sigma $--model action%
\begin{equation}
S=\int_{M}d^{2}\sigma \left( G_{MN}\,\eta _{\alpha \beta }\,\partial
^{\alpha }X^{M}\,\partial ^{\beta }X^{N}+B_{MN}\,\epsilon _{\alpha \beta
}\,\partial ^{\alpha }X^{M}\,\partial ^{\beta }X^{N}\right)
\end{equation}%
the background space-time metric is found as given in equ. (\ref{nw4_3}) and
the antisymmetric tensor field is read off to be $B_{12}=u=-B_{21}$ with all
the other components of $B_{MN}$ are zero. Then the magnetic field, $%
H_{u12}=\partial _{u}B_{12}$, is everywhere constant.

The generators of the isometry group of the background can be found by
noting that the WZNW action (\ref{WZWaction}) is invariant under the
transformation $g\rightarrow g_{L}\,g\,g_{R}$. Then the form of symmetry
generators are determined as%
\begin{equation}
\begin{array}{ll}
T=\partial _{v}\ , &  \\
J_{L}=\partial _{u}+(a_{1}\partial _{2}-a_{2}\partial _{1})\ ,\qquad &
J_{R}=\partial _{u}\ , \\
P_{L}^{1}=\partial _{1}+\frac{1}{2}a_{2}\partial _{v}\ , & P_{R}^{1}=\cos
u\,\partial _{1}+\sin u\,\partial _{2}+\frac{1}{2}\left( \sin u\,a_{1}-\cos
u\,a_{2}\right) \partial _{v}\ , \\
P_{L}^{2}=-\partial _{2}+\frac{1}{2}a_{1}\partial _{v}\ , & P_{R}^{2}=-\sin
u\,\partial _{1}+\cos u\,\partial _{2}+\frac{1}{2}\left( \cos u\,a_{1}+\sin
u\,a_{2}\right) \partial _{v}\ .%
\end{array}
\label{Killing}
\end{equation}

Therefore, the isometry group of NW spacetime is seven dimensional. It
contains two commuting, left and right, groups of isometry, which are both $%
H_{4}$. Since $T$ is the central element of the non-semi--simple algebra, it
is the same for left and right action. In particular, $T$ generates
translations in the $v$ direction, $J_{L}-J_{R}$ generates rotations in the $%
\left( a_{1},a_{2}\right) -$plane, and $J_{R}$ generates translations in the
$u$ direction. The remaining generators generate some \textquotedblleft
twisted translations\textquotedblright\ \cite{forgacs}.

The D'Alembertian in this background can be found as the Casimir of $H_{4}$
group by using the above forms of the symmetry generators (either left or
right ones). We will do this in subsection \ref{wavefunction} and then by
applying it on wave functional in a specific representation we are going to
derive the dispersion relation.

\subsection{Representations of the Non-semi--simple Algebra $h_{4}$\label%
{IrReps}}

In \ WZNW model, the spectrum of string theory on a specific group manifold
is built on the unitary irreducible representations of the corresponding Lie
algebra of the group. The irreducible representations of the Lie algebra
constitute the ground state level of the string. On this level all the
states have the same ground state energy. The excited states in higher
levels are then constructed by applying either the negative frequency modes
of the currents or the negative frequency modes of the free fields,
depending on the approach. Therefore the first step in constructing the
physical spectrum of string theory on a group manifold is to analyze the
unitary irreducible representations of the associated Lie algebra. In this
subsection we are going to analyze the unitary irreducible representations
of the non-semisimple algebra $h_{4}$.

The unitary irreducible representations of $h_{4}$ algebra are first
discussed in \cite{KK}. The authors of these papers noticed that the unitary
representations of $h_{4}$ algebra has very similar structure to the
representations of the $sl(2,R)$ algebra. However, they have \emph{missed}
one important type of unitary representations, inclusion of which makes the
similarity between unitary representations of $h_{4}$ and $sl(2,R)$ Lie
algebras truly remarkable.

To find the unitary irreducible representations we first rewrite the $h_{4}$
algebra in a different form as
\begin{equation}
\left[ P^{+},P^{-}\right] =-2iT,\quad \left[ J,P^{+}\right] =-iP^{+},\quad %
\left[ J,P^{-}\right] =iP^{-},\quad \left[ T,\ any\right] =0,  \label{h4_v2}
\end{equation}%
where $P^{+}=P^{1}+iP^{2}$ and $P^{-}=P^{1}-iP^{2}$. There are two Casimir
operators of this algebra: the central element $T$ and the quadratic
Casimir, which, in this basis for the algebra, has the form given by
\begin{equation}
C=\frac{1}{2}\left( P^{+}P^{-}+P^{-}P^{+}\right) +2JT\ .  \label{casimir}
\end{equation}%
The hermiticity properties of the generators are as follows%
\begin{equation}
\left( P^{+}\right) ^{\dag }=-P^{-}\ ,\quad J^{\dag }=-J\ ,\quad T^{\dag
}=-T\ .  \label{Herm_gen}
\end{equation}

Since the generator $T$ is an central element and, therefore, commutes with
every other member of the algebra, we can set it to a constant value
\begin{equation}
T\left\vert n,\vec{p}\right\rangle =-ip_{u}\left\vert n,\vec{p}\right\rangle
,  \label{operator_T}
\end{equation}%
where $\vec{p}=\left( p_{u}\,,p_{v}\,,p^{+},p^{-}\right) $ is a vector of
parameters. By notating a state of the representations like this we
anticipate the possible parameters that will appear in specific
representations. These parameters will turn out to be the same as the
independent momenta components in NW$_{4}$ spacetime. Therefore, the states $%
\left\vert n,\vec{p}\right\rangle $ in this basis can either be treated as
states in a \textquotedblleft number\textquotedblright\ basis, as it will be
seen in a moment, or states in a momentum basis.

To find the number basis consider an eigenstate of the Cartan generator,
\begin{equation}
J\,\left\vert 0,\vec{p}\right\rangle =-ip_{v}\,\left\vert 0,\vec{p}%
\right\rangle .
\end{equation}%
From the algebra it can easily be deduced that $P^{+}$ behaves as an
annihilation operator and $P^{-}$ behaves as a creation operator in a number
basis, in which one of the elements is the state $\left\vert 0,\vec{p}%
\right\rangle $. The $P^{+}$ and $P^{-}$ generators act on $\left\vert 0,%
\vec{p}\right\rangle $ as%
\begin{equation}
P^{\pm }\ \left\vert 0,\vec{p}\right\rangle =\sqrt{C+2p_{u}\,p_{v}\pm p_{u}}%
\ \left\vert \mp 1,\vec{p}\right\rangle \,,
\end{equation}%
where $C$ is the specific value of the quadratic Casimir operator in the
particular representation. According to different values of $C$ we obtained
the following classification of the representations of the Lie algebra $%
h_{4} $.

\textbf{[1]} Lowest--weight representations $\left( V^{p_{u},p_{v}},\
p_{u}>0\right) $: In such representations of the algebra, there is a so
called lowest--weight state on which $P^{+}$ has zero eigenvalue: $%
P^{+}\left\vert 0,\vec{p}\right\rangle =0$. The eigenvalues of the
generators on a general state in these representations are
\begin{eqnarray}
T\ \left\vert n,\vec{p}\right\rangle &=&-ip_{u}\ \left\vert n,\vec{p}%
\right\rangle , \\
J\ \left\vert n,\vec{p}\right\rangle &=&i\left( -p_{v}+n\right) \ \left\vert
n,\vec{p}\right\rangle ,  \notag \\
P^{+}\ \left\vert n,\vec{p}\right\rangle &=&i\sqrt{2p_{u}n}\ \left\vert n-1,%
\vec{p}\right\rangle ,  \notag \\
P^{-}\ \left\vert n,\vec{p}\right\rangle &=&i\sqrt{2p_{u}\left( n+1\right) }%
\ \left\vert n+1,\vec{p}\right\rangle .  \notag
\end{eqnarray}%
The spectrum of $-iJ$ is $\{-p_{v}+n\},\ n\in
\mathbb{N}
$ and the quadratic Casimir operator has the eigenvalue
\begin{equation}
C=-2p_{u}p_{v}-p_{u}.
\end{equation}
These representations were called \textquotedblleft Type
III\textquotedblright\ in \cite{KK}.

\textbf{[2]} Highest--weight representations $\left( \tilde{V}%
^{p_{u},p_{v}},\ p_{u}<0\right) $: In these representations of the algebra,
there is a so called highest--weight state on which $P^{-}$ has zero
eigenvalue: $P^{-}\left\vert 0,\vec{p}\right\rangle =0$. The eigenvalues of
the generators on a general state in these representations are
\begin{eqnarray}
T\ \left\vert n,\vec{p}\right\rangle &=&-ip_{u}\ \left\vert n,\vec{p}%
\right\rangle , \\
J\ \left\vert n,\vec{p}\right\rangle &=&i\left( -p_{v}+n\right) \ \left\vert
n,\vec{p}\right\rangle ,  \notag \\
P^{+}\ \left\vert n,\vec{p}\right\rangle &=&i\sqrt{2p_{u}\left( n-1\right) }%
\ \left\vert n-1,\vec{p}\right\rangle ,  \notag \\
P^{-}\ \left\vert n,\vec{p}\right\rangle &=&i\sqrt{2p_{u}n}\ \left\vert n+1,%
\vec{p}\right\rangle .  \notag
\end{eqnarray}%
The spectrum of $-iJ$ is $\{-p_{v}-n\},\ n\in
\mathbb{N}
$ and the quadratic Casimir operator has the eigenvalue%
\begin{equation}
C=-2p_{u}p_{v}+p_{u}.
\end{equation}%
These representations were called \textquotedblleft Type
II\textquotedblright\ in \cite{KK}.

\newpage

\textbf{[3]} Continuous series representations $\left(
C_{p^{+},p^{-}}^{p_{u},p_{v}}\right) $: Other than highest-weight and
lowest-weight representations it is possible to construct a series of
representations parametrized by two numbers $p^{+}$ and$\ p^{-}$. Here these
parameters are complex and they are complex conjugates of each other. In the
past $p_{u}$ have always been set to zero when analyzing this kind of
representations \cite{KK}--\cite{KP}, however we found that there are also
the following continuous series representations of $h_{4}$ Lie algebra:
\begin{eqnarray}
T\ \left\vert n,\vec{p}\right\rangle &=&-ip_{u}\ \left\vert n,\vec{p}%
\right\rangle ,  \label{Rep3} \\
J\ \left\vert n,\vec{p}\right\rangle &=&i\left( -p_{v}+n\right) \ \left\vert
n,\vec{p}\right\rangle ,  \notag \\
P^{+}\ \left\vert n,\vec{p}\right\rangle &=&i\sqrt{p^{+}p^{-}+2p_{u}n}\
\left\vert n-1,\vec{p}\right\rangle ,  \notag \\
P^{-}\ \left\vert n,\vec{p}\right\rangle &=&i\sqrt{p^{+}p^{-}+2p_{u}\left(
n+1\right) }\ \left\vert n+1,\vec{p}\right\rangle .  \notag
\end{eqnarray}%
These representations are the true analog of the principal continuous series
of $sl(2,R)$.

The spectrum of $-iJ$ is $\{-p_{v}+n\},\ n\in
\mathbb{Z}
$ and the quadratic Casimir operator has the eigenvalue
\begin{equation}
C=-2p_{u}p_{v}-p^{+}p^{-}-p_{u}.
\end{equation}

There are special cases of these representations. One can set $%
p^{+}=p^{-}=\alpha $, therefore decreasing number of parameters to one. We
denote such special cases as $C_{\alpha }^{p_{u},p_{v}}$. One can also set $%
p_{u}=0$ together with $p^{+}=p^{-}=\alpha $ , in which case one obtains the
representations called \textquotedblleft type I\textquotedblright\ in \cite%
{KK} and denoted as $V_{\alpha }^{0,p_{v}}$ in \cite{cheung}. One also
observes that, after setting $p^{+}p^{-}=0$, the lowest--weight
representations, $V^{p_{u},p_{v}},\ p_{u}>0$, are special cases of
continuous series representations.

\textbf{[4]} Conjugate continuous series representations $\left( \tilde{C}%
_{p^{+},p^{-}}^{p_{u},p_{v}}\right) $: These are again two parameter
continuous series representations. The eigenvalues of the generators on a
general state in these representations are
\begin{eqnarray}
T\ \left\vert n,\vec{p}\right\rangle &=&-ip_{u}\ \left\vert n,\vec{p}%
\right\rangle , \\
J\ \left\vert n,\vec{p}\right\rangle &=&i\left( -p_{v}+n\right) \ \left\vert
n,\vec{p}\right\rangle ,  \notag \\
P^{+}\ \left\vert n,\vec{p}\right\rangle &=&i\sqrt{p^{+}p^{-}+2p_{u}\left(
n-1\right) }\ \left\vert n-1,\vec{p}\right\rangle ,  \notag \\
P^{-}\ \left\vert n,\vec{p}\right\rangle &=&i\sqrt{p^{+}p^{-}+2p_{u}n}\
\left\vert n+1,\vec{p}\right\rangle .  \notag
\end{eqnarray}%
Here again one can set $p^{+}$ and $p^{-}$ to special values and obtain
special cases of these representations. For example setting $p^{+}p^{-}=0$
one obtains the highest-weight representations, $\tilde{V}^{p_{u},p_{v}},\
p_{u}<0$.

In these representations the spectrum of $-iJ$ is again $\{-p_{v}+n\},\ n\in
\mathbb{Z}
$ and the quadratic Casimir operator has the eigenvalue%
\begin{equation}
C=-2p_{u}p_{v}-p^{+}p^{-}+p_{u}.
\end{equation}

\textbf{[5]} Trivial representation: Setting $p_{u}=0=p_{v}$ one obtains the
trivial representation $V^{0,0}$.

\subsection{Wave Functional and the Dispersion Relation\label{wavefunction}}

In this section we are going to define Klauder--Perelomov generalized
coherent states \cite{KlPe} for the continuous series representation. Then
we are going to find the wave functional as the matrix element of the group
element in the coherent state basis. The same analysis is performed for the
highest and lowest weight representations in \cite{cheung} by using
Barut--Girardello type coherent states \cite{BR}. We refer the reader to
\cite{cheung} for the form of wave functions in those representations.

Klauder--Perelomov generalized coherent states are constructed by utilizing
the dynamical symmetry group of the particular system. For the continuous
series representation of the $H_{4}$ group they are defined by%
\begin{equation}
\left\vert \lambda \right) =e^{\lambda P^{+}-\bar{\lambda}P^{-}}\left\vert 0,%
\vec{p}\right\rangle .  \label{KPcoherent}
\end{equation}%
The action of generators of $H_{4}$ group on $\left\vert \lambda \right) $
can be easily determined as%
\begin{eqnarray}
T\ \left\vert \lambda \right) &=&-ip_{u}\ \left\vert \lambda \right) \,, \\
J\ \left\vert \lambda \right) &=&-i\left( p_{v}+\lambda \partial _{\lambda }-%
\bar{\lambda}\partial _{\bar{\lambda}}\right) \ \left\vert \lambda \right)
\,,  \notag \\
P^{+}\ \left\vert \lambda \right) &=&\left( \partial _{\lambda }+\bar{\lambda%
}p_{u}\right) \ \left\vert \lambda \right) \,,  \notag \\
P^{-}\ \left\vert \lambda \right) &=&\left( -\partial _{\bar{\lambda}%
}+\lambda p_{u}\right) \ \left\vert \lambda \right) \,,  \notag
\end{eqnarray}%
and the inner product in coherent state basis is found to be%
\begin{equation}
\left( \kappa |\lambda \right) =e^{-p_{u}\left( \lambda \bar{\lambda}+\kappa
\bar{\kappa}+2\bar{\kappa}\lambda \right) }\,_{1}F_{1}\left( 1+\frac{%
p^{+}p^{-}}{2p_{u}},1;\,2p_{u}\left( \lambda +\kappa \right) \left( \bar{%
\lambda}+\bar{\kappa}\right) \right) \ ,
\end{equation}%
where $_{1}F_{1}\left( \alpha ,\gamma ;z\right) $ is the confluent
hypergeometric function.

Then the wave functional as the matrix element of the group element of $%
H_{4} $ is found to have the form%
\begin{eqnarray}
\Psi _{\kappa ,\,\lambda \,}^{p_{v},\,p_{u},\,p^{+},\,p^{-}}\left(
u,v,a^{+},a^{-}\right) &=&\left( \kappa \right\vert \,g\,\left\vert \lambda
\right)  \notag \\
&=&\left( \kappa \right\vert
\,e^{a^{+}P^{-}+a^{-}P^{+}}e^{uJ+vT}\,\left\vert \lambda \right)
\label{wfunction} \\
&=&e^{-i\left( up_{v}+vp_{u}\right) }e^{-p_{u}\left(
-a^{+}a^{-}-2a^{+}\Lambda +\Lambda \bar{\Lambda}+\kappa \bar{\kappa}+2\left(
\Lambda +a^{-}\right) \bar{\kappa}\right) }\,_{1}F_{1}\left( \alpha
,1;w\right) \ ,  \notag
\end{eqnarray}%
where $\Lambda =e^{-iu}\lambda ,\ \bar{\Lambda}=e^{iu}\bar{\lambda},\ \alpha
=1+\frac{p^{+}p^{-}}{2p_{u}}$ and $w=2p_{u}\left( \Lambda +\kappa
+a^{-}\right) \left( \bar{\Lambda}+\bar{\kappa}-a^{+}\right) $.

This wave functional should obey the wave equation $\left( \square
-m^{2}\right) \Psi =0$, where the D'Alembertian operator is given as the
Casimir operator of the left or the right isometry groups. The generators of
the isometry group are given in Equ. (\ref{Killing}). Using the left or
right generators one finds%
\begin{equation}
\square =2\partial _{v}\partial _{u}+\partial _{1}^{2}+\partial
_{2}^{2}+\left( a_{1}\partial _{2}-a_{2}\partial _{1}\right) \partial _{v}+%
\frac{1}{4}\left( a_{1}^{2}+a_{2}^{2}\right) \partial _{v}^{2}\ .
\label{DAlembertian}
\end{equation}%
This formula is given in $\left( a_{1},a_{2}\right) $ coordinate system.
Since the wave functional in coherent state basis is given in $\left(
a^{+},a^{-}\right) $ coordinate system, we translate the D'Alembertian
operator into $\left( a^{+},a^{-}\right) $ coordinate system by using $a^{+}=%
\frac{1}{2}\left( a_{1}+ia_{2}\right) $ and $a^{-}=\frac{1}{2}\left(
a_{1}-ia_{2}\right) $. We find%
\begin{equation}
\square =2\partial _{v}\partial _{u}+\partial \bar{\partial}+i\left(
a^{+}\partial _{+}-a^{-}\partial _{-}\right) \partial
_{v}+a^{+}a^{-}\partial _{v}^{2}\ .  \label{DAlembertian2}
\end{equation}%
An arbitrary wave functional $\Psi _{n_{L},n_{R}}$ can be obtained by acting
$n_{L}$ and $n_{R}$ times the raising (lowering) operators $P_{L}^{\mp
}=P_{L}^{1}\mp iP_{L}^{2}$ and $P_{R}^{\mp }=P_{R}^{1}\mp iP_{R}^{2}$,
respectively, on the wave functional $\Psi $ (\ref{wfunction}). Then the
action of $\left( \square -m^{2}\right) $ on an arbitrary wave functional $%
\Psi _{n_{L},n_{R}}$ gives the dispersion relation as%
\begin{equation}
2p_{u}p_{v}+p^{+}p^{-}+p_{u}\left( n_{L}+n_{R}+1\right) +m^{2}=0\ .
\label{Dispertion}
\end{equation}

\section{Covariant Quantization and the Physical Spectrum}

\subsection{Free Field Realization of $\hat{h}_{4}$ Current Algebra\label%
{realization}}

In this section we are going to realize the holomorphic part of the current
algebra of $H_{4}$ WZNW model in terms of four bosons, $\theta ^{+},\ \theta
^{-},\ \phi ,\ \varphi $. Likewise the anti-holomorphic currents are going
to be realized by another four bosons, $\bar{\theta}^{+},\ \bar{\theta}%
^{-},\ \bar{\phi},\ \bar{\varphi}$. In the following we will often omit the
analysis of anti-holomorphic part. In those instances it should be
understood that the anti-holomorphic part works the same way as holomorphic
part with essentially trivial modifications. We start by writing the group
element as%
\begin{equation}
\begin{array}{ll}
g=g_{L}(z)g_{R}(\bar{z}), & g_{L}(z)=e^{\theta ^{+}P^{-}}e^{\phi J+\varphi
T}e^{\theta ^{-}P^{+}} \\
& g_{R}(\bar{z})=e^{\bar{\theta}^{+}P^{-}}e^{\bar{\phi}J+\bar{\varphi}T}e^{%
\bar{\theta}^{-}P^{+}}%
\end{array}
\label{g}
\end{equation}%
in terms of holomorphic and anti-holomorphic fields. Here in terms of
worldsheet coordinates $z=e^{i(\tau +\sigma )}$ and $\bar{z}=e^{i(\tau
-\sigma )}$. Notice that $z$ and $\bar{z}$ are independent complex
variables. The symmetry currents of the WZNW model are given in terms of the
group element by the relations $\mathcal{J}_{L}\left( z\right) =g\partial
_{z}g^{-1}$ for the holomorphic currents and $\mathcal{J}_{R}\left( \bar{z}%
\right) =g^{-1}\partial _{\bar{z}}g$ for the anti-holomorphic currents. In
terms of special form of the group element $g(z,\bar{z})=g_{L}(z)g_{R}(\bar{z%
})$ the currents become%
\begin{equation}
\mathcal{J}_{L}\left( z\right) =g_{L}\partial g_{L}^{-1},\qquad \mathcal{J}%
_{R}\left( \bar{z}\right) =g_{R}^{-1}\bar{\partial}g_{R},  \label{Curr_1}
\end{equation}%
where $\partial $ and $\bar{\partial}$ stand for $\partial _{z}$ and $%
\partial _{\bar{z}}$, respectively. Then using the above realization of left
and right group elements we find the currents in terms of holomorphic and
anti-holomorphic fields as%
\begin{equation}
\begin{array}{ll}
T_{L}\left( z\right) =-\partial \phi , & T_{R}\left( \bar{z}\right) =\bar{%
\partial}\bar{\phi} \\
J_{L}\left( z\right) =-\partial \varphi -i:\theta ^{+}\vartheta ^{-}:, &
J_{R}\left( \bar{z}\right) =\bar{\partial}\bar{\varphi}+i:\bar{\theta}^{-}%
\bar{\vartheta}^{+}: \\
P_{L}^{+}\left( z\right) =-2\,\partial \theta ^{+}+2i\,\partial \phi
\,\theta ^{+},\quad & P_{R}^{+}\left( \bar{z}\right) =\bar{\vartheta}^{+} \\
P_{L}^{-}\left( z\right) =-\vartheta ^{-}, & P_{R}^{-}\left( \bar{z}\right)
=2\,\bar{\partial}\bar{\theta}^{-}-2i\,\bar{\partial}\bar{\phi}\,\bar{\theta}%
^{-}%
\end{array}
\label{Curr_2}
\end{equation}%
where $\vartheta ^{-}\left( z\right) =2\,\partial \theta ^{-}e^{-i\phi }$
and $\bar{\vartheta}^{+}\left( \bar{z}\right) =2\,\bar{\partial}\bar{\theta}%
^{+}e^{-i\bar{\phi}}$ are bosonic free fields. At this point we would like
to remind the reader that we did not call any of the bosonic fields in the
realization of the group element as \textit{free} fields. The free fields in
our construction are $\theta ^{+},\ \vartheta ^{-},\ \phi $ and $\varphi $
from the point of view of holomorphic currents and $\bar{\theta}^{-},\ \bar{%
\vartheta}^{+},\ \bar{\phi}$ and$\ \bar{\varphi}$ from the point of view of
anti-holomorphic currents. With these points of view, the fields $\theta
^{-} $ and $\bar{\theta}^{+}$ are composite fields and are given in terms of
free fields as%
\begin{eqnarray}
\theta ^{-}\left( z\right) &=&x_{0}^{-}+\frac{1}{2}\int^{z}dz^{^{\prime
}}\,\vartheta ^{-}(z^{^{\prime }})e^{i\phi (z^{^{\prime }})}\,,
\label{thetaM} \\
\bar{\theta}^{+}\left( \bar{z}\right) &=&\bar{x}_{0}^{+}+\frac{1}{2}\int^{%
\bar{z}}d\bar{z}^{^{\prime }}\bar{\vartheta}^{+}(\bar{z}^{^{\prime }})e^{i%
\bar{\phi}(\bar{z}^{^{\prime }})}\,.
\end{eqnarray}%
These expressions will be used when we discuss about the operator products
of vertex operator with the currents and with the energy-momentum tensor.

We should warn the reader that in the present realization the pairs of
fields $\theta ^{+},\ \vartheta ^{-}$ and $\bar{\theta}^{-},\ \bar{\vartheta}%
^{+}$ are not ghost fields, they have bosonic character. In fact the mode
expansions of all the holomorphic fields are as follows%
\begin{eqnarray}
\phi \left( z\right) &=&\phi _{0}+i\alpha _{0}^{-}\ln z-i\sum\limits_{n\neq
0}\frac{1}{n}\alpha _{n}^{-}z^{-n}  \label{phi} \\
\varphi \left( z\right) &=&\varphi _{0}-i\alpha _{0}^{+}\ln
z+i\sum\limits_{n\neq 0}\frac{1}{n}\alpha _{n}^{+}z^{-n}  \label{varphi} \\
\vartheta ^{-}\left( z\right) &=&i\sum\limits_{n=-\infty }^{\infty }\beta
_{n}^{-}z^{-n-1}  \label{vartheta} \\
\theta ^{+}\left( z\right) &=&x_{0}^{+}-i\beta _{0}^{+}\ln
z+i\sum\limits_{n\neq 0}\frac{1}{n}\beta _{n}^{+}z^{-n}  \label{theta}
\end{eqnarray}%
The anti-holomorphic free fields have similar mode expansions. The free
fields $\phi \left( z\right) $ and $\varphi \left( z\right) $ are Hermitian,
whereas $\vartheta ^{-}\left( z\right) $ and $\partial \theta ^{+}\left(
z\right) $ are Hermitian conjugates of each other:
\begin{equation}
\left[ \phi \left( z\right) \right] ^{\dag }=\phi \left( \frac{1}{z}\right)
,\quad \left[ \varphi \left( z\right) \right] ^{\dag }=\varphi \left( \frac{1%
}{z}\right) ,\quad \left[ z\vartheta ^{-}\left( z\right) \right] ^{\dag }=%
\frac{1}{z}\partial \theta ^{+}\left( \frac{1}{z}\right) .
\label{Hermiticity}
\end{equation}%
Then the Hermicity of the modes follows as%
\begin{equation}
\left( \alpha _{n}^{-}\right) ^{\dag }=\alpha _{-n}^{-}\ ,\quad \left(
\alpha _{n}^{+}\right) ^{\dag }=\alpha _{-n}^{+}\ ,\quad \left( \beta
_{n}^{-}\right) ^{\dag }=\beta _{-n}^{+}\ ,\quad \left( \beta
_{n}^{+}\right) ^{\dag }=\beta _{-n}^{-}\ .\quad  \label{Herm_modes}
\end{equation}%
We emphasize that under Hermitian conjugation the $\alpha ^{-}$ and $\alpha
^{+}$ modes are conjugate to their kinds, however the $\beta ^{-}$ and $%
\beta ^{+}$ modes switch to each other.

Non-zero frequency modes of the free fields have the following commutation
relations with each other:%
\begin{equation}
\left[ \alpha _{n}^{-},\alpha _{m}^{+}\right] =n\delta _{n+m},\qquad \left[
\beta _{n}^{-},\beta _{m}^{+}\right] =n\delta _{n+m}.  \label{Comm_n}
\end{equation}%
We also choose the following commutation relations between the zero modes of
the free fields:%
\begin{equation}
\left[ \phi _{0},\alpha _{0}^{+}\right] =-i,\qquad \left[ \varphi
_{0},\alpha _{0}^{-}\right] =i,\qquad \left[ x_{0}^{+},\beta _{0}^{-}\right]
=i.  \label{Comm_0}
\end{equation}%
These zero modes are normal ordered as%
\begin{eqnarray}
\colon \beta _{0}^{-}x_{0}^{+}\colon &=&x_{0}^{+}\beta _{0}^{-}\quad
\Longrightarrow \quad \left\langle \beta _{0}^{-}x_{0}^{+}\right\rangle
=-i\quad ,\quad \colon x_{0}^{+}\beta _{0}^{-}\colon =x_{0}^{+}\beta
_{0}^{-}\quad \Longrightarrow \quad \left\langle x_{0}^{+}\beta
_{0}^{-}\right\rangle =0,  \notag \\
\colon \alpha _{0}^{-}\varphi _{0}\colon &=&\varphi _{0}\alpha _{0}^{-}\quad
\Longrightarrow \quad \left\langle \alpha _{0}^{-}\varphi _{0}\right\rangle
=-i\quad ,\quad \colon \varphi _{0}\alpha _{0}^{-}\colon =\varphi _{0}\alpha
_{0}^{-}\quad \Longrightarrow \quad \left\langle \varphi _{0}\alpha
_{0}^{-}\right\rangle =0,  \notag \\
\colon \alpha _{0}^{+}\phi _{0}\colon &=&\phi _{0}\alpha _{0}^{+}\quad
\Longrightarrow \quad \left\langle \alpha _{0}^{+}\phi _{0}\right\rangle
=i\quad ,\quad \colon \phi _{0}\alpha _{0}^{+}\colon =\phi _{0}\alpha
_{0}^{+}\quad \Longrightarrow \quad \left\langle \phi _{0}\alpha
_{0}^{+}\right\rangle =0.  \label{zeroOrder}
\end{eqnarray}

Then the contractions of the bosonic free fields are calculated to be equal
to%
\begin{eqnarray}
\left\langle \phi \left( z\right) \varphi \left( w\right) \right\rangle
&=&\ln \left( z-w\right) ,  \label{fifi} \\
\left\langle \varphi \left( z\right) \phi \left( w\right) \right\rangle
&=&\ln \left( z-w\right) , \\
\left\langle \vartheta ^{-}\left( z\right) \theta ^{+}\left( w\right)
\right\rangle &=&\frac{1}{z-w}, \\
\left\langle \theta ^{+}\left( z\right) \vartheta ^{-}\left( w\right)
\right\rangle &=&-\frac{1}{z-w}.  \label{tete}
\end{eqnarray}

Now we can calculate the operator products among currents and check that we
have a correct realization of the currents in terms of free fields. The
operator products come out to be the correct ones:%
\begin{eqnarray}
T_{L}\left( z\right) J_{L}\left( w\right) &\sim &\frac{1}{\left( z-w\right)
^{2}}  \label{Contr_Curr} \\
J_{L}\left( z\right) P_{L}^{+}\left( w\right) &\sim &\frac{-i}{z-w}%
P_{L}^{+}\left( w\right)  \notag \\
J_{L}\left( z\right) P_{L}^{-}\left( w\right) &\sim &\frac{i}{z-w}%
P_{L}^{-}\left( w\right)  \notag \\
P_{L}^{+}\left( z\right) P_{L}^{-}\left( w\right) &\sim &\frac{2}{\left(
z-w\right) ^{2}}-\frac{2i}{z-w}T_{L}\left( w\right)  \notag
\end{eqnarray}%
These are of course the singular parts of the operator products.

Next we calculate the energy--momentum tensor. In terms of the affine
currents the energy--momentum tensor is defined in Sugawara form as%
\begin{equation}
\mathcal{T}_{L}\left( z\right) =\frac{1}{2}L_{ij}:\mathcal{J}_{L}^{i}%
\mathcal{J}_{L}^{j}:,  \label{EM_def}
\end{equation}%
where sum over $i$ and $j$ indices is implicit and $:\ :$ means that one
should subtract any singular part from the OPE of current--current products.
We determine the form of $L_{ij}$ by requiring the affine symmetry currents
to be primary with respect to the energy--momentum tensor. This means that
the currents should have conformal weight one and, therefore, their OPE with
the energy--momentum tensor should have the form%
\begin{equation}
\mathcal{T}_{L}\left( z\right) \,\mathcal{J}_{L}^{i}(w)=\frac{1}{\left(
z-w\right) ^{2}}\mathcal{J}_{L}^{i}(w)+\frac{1}{z-w}\partial _{w}\mathcal{J}%
_{L}^{i}(w)
\end{equation}%
If we did not required this condition, then energy--momentum tensor would
have to obey a Virasoro master equation \cite{HK},\cite{KK},\cite{Moh93}. We
find that the currents are primary if the matrix $L_{ij}$ is the same as the
non-degenerate bilinear form $\Omega _{ij}$ for the group $H_{4}$. Then the
holomorphic energy--momentum tensor is calculated as%
\begin{eqnarray}
\mathcal{T}_{L}\left( z\right) &=&\frac{1}{2}\colon \left\{ T_{L}\left(
z\right) J_{L}\left( z\right) +J_{L}\left( z\right) T_{L}\left( z\right) +%
\frac{1}{2}\left[ P_{L}^{+}\left( z\right) P_{L}^{-}\left( z\right)
+P_{L}^{-}\left( z\right) P_{L}^{+}\left( z\right) \right] \right\} \colon
\notag \\
&=&\colon \partial \phi \left( z\right) \partial \varphi \left( z\right)
\colon +\colon \partial \theta ^{+}\left( z\right) \vartheta ^{-}\left(
z\right) \colon +\frac{i}{2}\partial ^{2}\phi \left( z\right)
\label{EM_holo}
\end{eqnarray}

In the CFT analysis of the spectrum of string theory, the zero mode in the
Laurent expansion of the holomorphic energy--momentum tensor plays the role
of one of the parts of the Hamiltonian. The other part comes symmetrically
from the anti-holomorphic energy--momentum tensor. To find the so called
Virasoro generators we expand $\mathcal{T}_{L}\left( z\right) $ as%
\begin{equation}
\mathcal{T}_{L}\left( z\right) =\sum\limits_{n=-\infty }^{\infty
}L_{n}\,z^{-n-2}.  \label{EM_exp}
\end{equation}%
Then using the mode expansions of the free fields the forms of $L_{n}$ and $%
L_{0}$ are determined as%
\begin{eqnarray}
L_{n} &=&\sum\limits_{m=-\infty }^{\infty }\left( \colon \alpha
_{n-m}^{-}\alpha _{m}^{+}\colon +\colon \beta _{n-m}^{-}\beta _{m}^{+}\colon
\right) +\frac{\alpha _{n}^{-}}{2}\ ,  \label{Ln} \\
L_{0} &=&\frac{\alpha _{0}^{-}}{2}+\alpha _{0}^{-}\alpha _{0}^{+}+\beta
_{0}^{-}\beta _{0}^{+}+\sum\limits_{m>0}\left( \alpha _{-m}^{+}\alpha
_{m}^{-}+\alpha _{-m}^{-}\alpha _{m}^{+}+\beta _{-m}^{+}\beta _{m}^{-}+\beta
_{-m}^{-}\beta _{m}^{+}\right) \ .  \label{L0}
\end{eqnarray}%
$L_{n}$'s obey Virasoro algebra with central charge equal to $4$.
Interestingly, this charge, at the same time, is equal to the dimension of
the $H_{4}$ manifold and is independent of the Kac--Moody level $k$, unlike
in the cases of WZNW\ models on semi-simple group manifolds. To make the
bosonic string theory on $H_{4}$ manifold critical, one also needs an
internal CFT with $c=22$. We are not going to tell anything more about the
internal CFT in this paper. We will determine the spectrum without referring
to any special properties of such internal CFT.

\subsection{Virasoro Constraints and the Physical Spectrum\label{Spectrum}}

In the previous section we have successfully mapped the affine currents to a
system of four free fields. Then, in this section, we determine the spectrum
of bosonic strings on NW spacetime as a Fock space created by the negative
frequency modes of four bosonic free fields as%
\begin{equation}
\prod\limits_{i,\,j,\,k,\,l=1}^{\infty }\left( \alpha _{-i}^{-}\right)
^{e_{1,i}}\left( \alpha _{-j}^{+}\right) ^{e_{2,j}}\left( \beta
_{-k}^{-}\right) ^{e_{3,k}}\left( \beta _{-l}^{+}\right) ^{e_{4,l}}\
\left\vert n;p_{u}\,,p_{v}\,,p^{+},p^{-}\right\rangle
\end{equation}%
where $\left\vert n;p_{u}\,,p_{v}\,,p^{+},p^{-}\right\rangle \equiv
\left\vert n,\vec{p}\right\rangle $ are the states in the base level and the
powers $e_{\mu ,i}\ (\mu =1,\ldots ,4)$ are positive integers or zero. This
Fock space is equivalent to the continuous series representation of the
Kac--Moody algebra $\hat{h}_{4}$. Since we have four free bosons, this
spectrum is no different than the spectrum of string theory in $4$
dimensional flat space-time. States created by $\beta _{-m}^{-}\,,\ \beta
_{-m}^{+}$ and $\alpha _{-m}^{1}=\frac{1}{\sqrt{2}}\left( \alpha
_{-m}^{+}+\alpha _{-m}^{-}\right) $ have positive norms, but there are
ghosts in this spectrum due to time-like oscillator modes $\alpha _{-m}^{0}=%
\frac{1}{\sqrt{2}}\left( \alpha _{-m}^{+}-\alpha _{-m}^{-}\right) $.
However, the situation is no worse than the case in string theory in flat
space-time. Here, again, we expect that the Virasoro constraints,%
\begin{equation}
\left( L_{n}-a\delta _{n}\right) \,\left\vert phys\right\rangle =0\quad
\left( n\geq 0\right) ,  \label{physical}
\end{equation}%
where $a=1$ in the critical case, will be enough to clean the spectrum from
unwanted negative-norm states.

We first look at the meaning of Virasoro constraints on the ground level
states in the critical case. Since the positive frequency modes of free
fields annihilate these states, we find that the eigenvalues of $L_{m}\
\left( m\geq 0\right) $ operators are
\begin{eqnarray}
L_{m}\left\vert n,\vec{p}\right\rangle &=&0\quad \left( m>0\right) , \\
L_{0}\left\vert n,\vec{p}\right\rangle &=&\left( \alpha _{0}^{-}\alpha
_{0}^{+}+\beta _{0}^{-}\beta _{0}^{+}+\frac{\alpha _{0}^{-}}{2}\right)
\left\vert n,\vec{p}\right\rangle =\left( p_{u}p_{v}+\frac{p^{-}p^{+}}{2}+%
\frac{p_{u}}{2}\right) \left\vert n,\vec{p}\right\rangle =\left\vert n,\vec{p%
}\right\rangle .  \notag \\
&&  \label{L0_ground}
\end{eqnarray}%
Due to the constraint $L_{n}=\bar{L}_{n}$, the anti-holomorphic Virasoro
generators have the same eigenvalues on the ground level states. Then, since
the covariant Hamiltonian of bosonic string theory is given by the sum $%
H=L_{0}+\bar{L}_{0}$ of holomorphic and anti-holomorphic zero frequency
Virasoro generators, the eigenvalue of the Hamiltonian on ground level
states is found to be the same as the dispersion relation derived in
subsection \ref{wavefunction}. Therefore all the states in the ground level
have the same energy. This is also true for the higher levels: All the
states in the same level of continuous series representation of $\hat{h}_{4}$
Kac--Moody algebra have the same energy. This will cause the same
regularization problem in the one-loop partition function of closed strings
as in the SL(2,R) WZNW model.

From the constraint (\ref{L0_ground}) we read the eigenvalues of the zero
modes on ground level states as%
\begin{eqnarray}
\alpha _{0}^{-}\,\left\vert n,\vec{p}\right\rangle &=&p_{u}\left\vert n,\vec{%
p}\right\rangle ,\quad \alpha _{0}^{+}\,\left\vert n,\vec{p}\right\rangle
=p_{v}\left\vert n,\vec{p}\right\rangle , \\
\beta _{0}^{-}\,\left\vert n,\vec{p}\right\rangle &=&\frac{p^{-}}{\sqrt{2}}%
\left\vert n,\vec{p}\right\rangle ,\quad \beta _{0}^{+}\,\left\vert n,\vec{p}%
\right\rangle =\frac{p^{+}}{\sqrt{2}}\left\vert n,\vec{p}\right\rangle .\quad
\end{eqnarray}%
At a higher level $l$ the $\left( L_{0}-a\right) \,\left\vert
phys\right\rangle =0$\ constraint gives the mass shell condition as%
\begin{equation}
p_{u}p_{v}+\frac{p^{+}p^{-}}{2}+\frac{p_{u}}{2}+l=a
\end{equation}

Next, continuing \`{a} la Bars \cite{B95} we determine the physical states
at level 1. We find the following combinations of states of Fock space obey
the physical state conditions (\ref{physical}):
\begin{eqnarray}
\left\vert \Phi _{\alpha }\right\rangle &=&\left[ \left( p_{v}+\frac{1}{2}%
\right) \alpha _{-1}^{-}-p_{u}\alpha _{-1}^{+}\right] \left\vert n,\vec{p}%
\right\rangle \ ,  \notag \\
\left\vert \Phi _{\beta }\right\rangle &=&\frac{1}{\sqrt{2}}\left(
p^{+}\beta _{-1}^{-}-p^{-}\beta _{-1}^{+}\right) \left\vert n,\vec{p}%
\right\rangle \ ,  \label{DDF} \\
\left\vert \Phi _{m}\right\rangle &=&\left[ p^{+}p^{-}\left( p_{u}\alpha
_{-1}^{+}+\left( p_{v}+\frac{1}{2}\right) \alpha _{-1}^{-}\right) -\frac{%
2p_{u}p_{v}+p_{u}}{\sqrt{2}}\left( p^{+}\beta _{-1}^{-}+p^{-}\beta
_{-1}^{+}\right) \right] \left\vert n,\vec{p}\right\rangle  \notag
\end{eqnarray}%
These states are orthogonal to each other and we found the norms of these
states as
\begin{eqnarray}
\left\langle \Phi _{\alpha }|\Phi _{\alpha }\right\rangle &=&\left(
2p_{u}p_{v}-p_{u}\right) =p^{+}p^{-}+2\left( 1-a\right) >0,  \notag \\
\left\langle \Phi _{\beta }|\Phi _{\beta }\right\rangle &=&\,p^{+}p^{-}>0,
\notag \\
\left\langle \Phi _{m}|\Phi _{m}\right\rangle &=&\,2p^{+}p^{-}\left(
p^{+}p^{-}+2\left( 1-a\right) \right) \left( 1-a\right) ,  \label{Norms}
\end{eqnarray}%
where we used the mass-shell condition at level 1,%
\begin{equation}
p_{u}p_{v}+\frac{p^{+}p^{-}}{2}+\frac{p_{u}}{2}+1=a
\end{equation}%
to obtain the the last relation in (\ref{Norms}). As it is seen, the norms
of the first two states in (\ref{Norms}) are positive. Whereas, the norm of
the last state in (\ref{Norms}) is positive only for $a<1$. It becomes a
null state in the critical case, $a=1$, and a ghost for $a>1$. This is the
same situation as in the $4$ dimensional flat bosonic string theory if we
take it as a part of string theory in $26$ dimensions. We are encountering
the same kind of situation here, because we mapped our theory to the theory
of four bosonic free fields as in the case of $4$ dimensional flat string
theory.

The physical state combinations at higher levels can be constructed,
therefore, just as in the case of $4$ dimensional flat string theory. Thus
there is no need to repeat that construction here. It is possible to prove
the non-existence of negative norm states in the spectrum just as in the
flat string theory \cite{thorn}. We would like to emphasize that the
spectrum we are proposing is free of ghosts without the requirement of an
extra constraint. Such a constraint, which is $p_{u}<1$, is mentioned first
in an unpublished paper by Forgacs et al. \cite{macars}, and later again in
\cite{forgacs},\cite{KA},\cite{cheung}. Since it is very similar to the
analogous constraint in the case of SL(2,R) WZNW\ model, it seemed as a
plausible constraint. However, in the next short subsection we comment on
the origin of this constraint. We are going to claim that the $p_{u}<1$
condition is an artifact of the formalism previously used.

\subsubsection{Comments on $p_{u}<1$ Condition}

To understand the $p_{u}<1$ condition we map the theory into $2$ bosonic
free fields and $2$ bosonic ghosts as in \cite{cheung}. Ghosts, together,
construct a $\beta -\gamma $ system. Then the currents in terms of these
fields are given as%
\begin{eqnarray}
T_{L}\left( z\right) &=&-\partial \phi \ ,  \notag \\
J_{L}\left( z\right) &=&-\partial \varphi -i:\gamma \beta :\ ,  \notag \\
P_{L}^{+}\left( z\right) &=&-2\,\partial \gamma +2i\,\partial \phi \,\gamma
\ ,  \notag \\
P_{L}^{-}\left( z\right) &=&-\beta \ ,
\end{eqnarray}%
where the bosonic ghost fields, $\beta $ and $\gamma $, have the mode
expansions:
\begin{equation}
\beta \left( z\right) =\sum_{n}\beta _{n}z^{-n-1}\,,\qquad \gamma \left(
z\right) =\sum_{n}\gamma _{n}z^{-n}
\end{equation}%
with commutation relation among the modes being $\left[ \beta _{n},\gamma
_{m}\right] =\delta _{n+m}$. The commutation relations of the modes of the
bosonic free fields are as in subsection \ref{realization}. The modes of
currents in terms of modes of the bosonic and the ghost fields can be easily
determined as%
\begin{eqnarray}
T_{n} &=&-i\alpha _{n}^{-}\,,  \notag \\
J_{n} &=&i\alpha _{n}^{+}-i\sum_{m}:\beta _{n-m}\gamma _{m}:\,,  \notag \\
P_{n}^{+} &=&2n\gamma _{n}-2\sum_{m}\alpha _{n-m}^{-}\gamma _{m}\,,  \notag
\\
P_{n}^{-} &=&-\beta _{n}
\end{eqnarray}%
The Hermiticity properties of the modes of currents can be deduced from the
Hermiticity properties of the zero frequency modes of currents, which are
the generators of the Lie algebra $h_{4}$. Then the Hermiticity properties
of the modes are
\begin{equation}
\left( T_{n}\right) ^{\dag }=-T_{-n}\,,\quad \left( J_{n}\right) ^{\dag
}=-J_{-n}\,,\quad \left( P_{n}^{+}\right) ^{\dag }=P_{-n}^{-}\,,\quad \left(
P_{n}^{-}\right) ^{\dag }=P_{-n}^{+}\,,\
\end{equation}

Now, consider the excited state $\left\vert \Phi \right\rangle
=P_{-1}^{+}\left\vert base\right\rangle $. Its norm is%
\begin{equation}
\left\langle \Phi |\Phi \right\rangle =\left( 2-2p_{u}\right) .
\end{equation}%
In the calculation we used the fact that $\alpha _{0}^{-}\left\vert
base\right\rangle =iT_{0}\left\vert base\right\rangle =-p_{u}\left\vert
base\right\rangle $ \cite{cheung}. Therefore, for the norm of the state $%
\left\vert \Phi \right\rangle $ to be positive, $p_{u}<1$ condition is
required. Thus this condition is an artifact of the formalism in which the
currents are realized in terms of bosonic and ghost fields, and the states
are constructed by using the modes of currents, but not the modes of the set
of free fields, to which the currents are mapped.

\subsection{Monodromy\label{monod}}

Since we realize the affine currents of $H_{4}$ WZNW model in terms of
bosonic free fields, the currents, when expanded in terms of modes of these
bosonic free fields, contain logarithmic cuts and consequently not analytic.
We recall that the equations of motion derived from WZNW action \cite{GW86},%
\begin{equation}
\partial _{\bar{z}}\,J_{L}=0\,,\qquad \partial _{z}\,J_{R}=0
\end{equation}%
require only that the currents separate into the holomorphic and the
anti-holomorphic parts. There are no constraints that require the currents
to be analytic form the start. However, their matrix elements in the
physical sector of the spectrum should be analytic, in fact periodic. This
is due to the boundary condition in the closed string sector, which require
that the fields on the string worldsheet to be periodic in the $\sigma $
variable. The corresponding requirement on the radial quantization is that
the fields should be invariant under the monodromy transformations as $%
z\rightarrow ze^{i2\pi n}$. That is instead of taking the currents as
periodic from the beginning, the periodicity is implemented in the Hilbert
space \cite{B95}. Since the energy-momentum tensor (\ref{EM_holo}) turned
out to be analytic and therefore it was possible to mode expand it in terms
of Virasoro generators, the monodromy operation commutes with the Virasoro
generators and can be implemented in Hilbert space without any problems.

The monodromy condition on the currents means that%
\begin{equation}
\left\langle phys^{\prime }\right\vert \,\mathcal{J}_{L}\left( ze^{i2\pi
n}\right) \,\left\vert phys\right\rangle =\left\langle phys^{\prime
}\right\vert \,\mathcal{J}_{L}\left( z\right) \,\left\vert phys\right\rangle
\label{monodromy}
\end{equation}%
From the form of the currents in terms of free fields (\ref{Curr_2}) one
finds that the currents undergo a linear transformation under the monodromy:%
\begin{eqnarray}
T_{L}\left( ze^{i2\pi n}\right) &=&T_{L}\left( z\right) \ ,  \notag \\
J_{L}\left( ze^{i2\pi n}\right) &=&J_{L}\left( z\right) +i2\pi n\,\beta
_{0}^{+}\,P_{L}^{-}\left( z\right) \ ,  \notag \\
P_{L}^{+}\left( ze^{i2\pi n}\right) &=&P_{L}^{+}\left( z\right) -i4\pi
n\,\beta _{0}^{+}\,T_{L}\left( z\right) \ ,  \notag \\
P_{L}^{-}\left( ze^{i2\pi n}\right) &=&P_{L}^{-}\left( z\right) \ .
\end{eqnarray}%
Since this transformation is linear, it can be written as an adjoint action
and since $P_{L}^{-}\left( z\right) $ does not change under monodromy, its
zero mode should be the generator of the transformation:%
\begin{equation}
\mathcal{J}_{L}\left( ze^{i2\pi n}\right) =e^{-i2\pi n\beta _{0}^{+}\beta
_{0}^{-}}\,\mathcal{J}_{L}\left( z\right) \,e^{i2\pi n\beta _{0}^{+}\beta
_{0}^{-}}\,,
\end{equation}%
where $\beta _{0}^{+}$ just acts as a number and $-i\beta _{0}^{-}$ is the
zero mode of $P_{L}^{-}\left( z\right) $. Then, physical states which
satisfy the monodromy condition (\ref{monodromy}) are the states that are
invariant under the monodromy transformation%
\begin{equation}
e^{-i2\pi n\beta _{0}^{+}\beta _{0}^{-}}\,\left\vert phys\right\rangle
=\left\vert phys\right\rangle
\end{equation}

Implementing this condition on the states in the Fock space%
\begin{equation}
e^{-i2\pi n\beta _{0}^{+}\beta
_{0}^{-}}\prod\limits_{i,\,j,\,k,\,l=1}^{\infty }\left( \alpha
_{-i}^{-}\right) ^{e_{1,i}}\left( \alpha _{-j}^{+}\right) ^{e_{2,j}}\left(
\beta _{-k}^{-}\right) ^{e_{3,k}}\left( \beta _{-l}^{+}\right) ^{e_{4,l}}\
\left\vert n;p_{u}\,,p_{v}\,,p^{+},p^{-}\right\rangle
\end{equation}%
one finds that the monodromy condition does not affect the form of the
physical states at the higher levels, but the product of momenta in
transverse coordinates is quantized in terms of non-negative integers:
\begin{equation}
e^{-i2\pi n\beta _{0}^{+}\beta _{0}^{-}}\left\vert n,\vec{p}\right\rangle
=\left\vert n,\vec{p}\right\rangle \quad \Longrightarrow \quad \beta
_{0}^{+}\beta _{0}^{-}=\frac{1}{2}p^{+}p^{-}=r\geq 0.\
\end{equation}%
Here the reason we take $r$ to be a non-negative integer is that according
to mass-shell condition the $p^{+}p^{-}$ product is non-negative. Then, the
mass-shell condition at excitation level $l$ becomes%
\begin{equation}
p_{u}p_{v}+r+\frac{p_{u}}{2}+l=a\,.
\end{equation}

\section{Vertex Operator}

In the previous section, we have shown that the physical spectrum of the
quantum string on NW spacetime is the affine continuous series
representation of the Kac--Moody algebra $\hat{h}_{4}$. According to the
\textit{state-operator correspondence} hypothesis, to each state in the
physical spectrum, there corresponds a vertex operator. In this section, we
are going to write the most basic vertex operator that corresponds to the $%
\left\vert 0,\vec{p}\right\rangle $ state in ground level of the string
spectrum. This \textquotedblleft tachyon\textquotedblright\ vertex operator
can be written as the group element in the continuous series representation
of $\hat{h}_{4}$ as%
\begin{equation}
V\left( g\right) =e^{a^{+}\hat{P}^{-}+a^{-}\hat{P}^{+}}e^{u\hat{J}+v\hat{T}%
}\,,
\end{equation}%
or from the group property, in any representation we can write $V\left(
g\right) =V\left( g_{L}\right) V\left( g_{R}\right) $ with%
\begin{equation}
V\left( g_{L}\right) =e^{\theta ^{+}\hat{P}^{-}}e^{\phi \hat{J}+\varphi \hat{%
T}}e^{\theta ^{-}\hat{P}^{+}}\,,\qquad V\left( g_{R}\right) =e^{\bar{\theta}%
^{+}\hat{P}^{-}}e^{\bar{\phi}\hat{J}+\bar{\varphi}\hat{T}}e^{\bar{\theta}^{-}%
\hat{P}^{+}}\,,
\end{equation}%
where $V\left( g_{L}\right) $ and $V\left( g_{R}\right) $ are constructed
from free fields of previous section. Here $\hat{T},\ \hat{J},\ \hat{P}^{+}$
and $\hat{P}^{-}$ are some operator representations of the generators of $%
h_{4}$. We found that the following position-momentum basis representation
of generators correspond to the continuous series representation of the
algebra $h_{4}$:%
\begin{equation}
\begin{array}{lll}
\hat{T} & \equiv -ip_{u} &  \\
\hat{J} & \equiv -ip_{v}+\hat{p}^{-}\hat{x}^{+} & \equiv -ip_{v}-\hat{x}^{-}%
\hat{p}^{+} \\
\hat{P}^{+} & \equiv 2p_{u}\,\hat{x}^{+}+2i\frac{r}{\hat{p}^{-}}\quad &
\equiv i\hat{p}^{+} \\
\hat{P}^{-} & \equiv i\hat{p}^{-} & \equiv -2p_{u}\,\hat{x}^{-}+2i\frac{r}{%
\hat{p}^{+}}%
\end{array}
\label{operGen}
\end{equation}%
In these representations of the generators, $\left( \hat{x}^{+},\hat{p}%
^{-}\right) $ and $\left( \hat{x}^{-},\hat{p}^{+}\right) $\ are canonically
conjugate pairs. $r$ is a quantum number of the states in the continuous
series representation of $h_{4}$ as determined by the monodromy
considerations. It should be noted that in the representation on the left
there is only the $\left( \hat{x}^{+},\hat{p}^{-}\right) $ conjugate pair,
and in the representation on the right there is only the $\left( \hat{x}^{-},%
\hat{p}^{+}\right) $ conjugate pair. In order to prevent possible confusion
we remark that these $\hat{x}^{+},\ \hat{p}^{-},\ etc.$ operators should not
be thought as the zero modes of the free fields (\ref{phi}-\ref{theta}).
They are just a convenient device for a representation of generators in the
continuous series representation of $h_{4}$. Using the canonical commutation
relations $\left[ \hat{x}^{+},\hat{p}^{-}\right] =i=\left[ \hat{x}^{-},\hat{p%
}^{+}\right] $ it can easily to be shown that $\hat{T},\ \hat{J},\ \hat{P}%
^{+}$ and $\hat{P}^{-}$ obey correct commutation relations of the algebra $%
h_{4}$. To be sure that these operator representations really correspond to
the continuous series representation of $h_{4}$ we also calculate the
quadratic Casimir operator. For either form of the generators it is%
\begin{eqnarray}
\hat{C} &=&\frac{1}{2}\left( \hat{P}^{+}\hat{P}^{-}+\hat{P}^{-}\hat{P}%
^{+}\right) +\hat{J}\,\hat{T}+\hat{T}\,\hat{J}  \notag \\
&=&-2p_{u}p_{v}-2r-p_{u}.
\end{eqnarray}%
Recall that in subsection \ref{monod}, from monodromy considerations we have
found that $r=\frac{1}{2}p^{+}p^{-}$. Therefore this representation of
generators correspond to the continuous series representation of $h_{4}$.

We now consider states on which either $\hat{P}^{+}$ or $\hat{P}^{-}$
operators are diagonal. These states are the momentum basis states $%
\left\langle p^{-},p_{u},p_{v}\right\vert \equiv \left\langle
p^{-}\right\vert $, on which $\hat{P}^{-}\equiv i\hat{p}^{-}$ is diagonal,
and $\left\vert p^{+},p_{u},p_{v}\right\rangle \equiv \,\left\vert
p^{+}\right\rangle $, on which $\hat{P}^{+}\equiv i\hat{p}^{+}$ is diagonal.
In the momentum basis we have $\left\langle p^{-}\right\vert \,\hat{x}^{+}=i%
\frac{\partial }{\partial p^{-}}\left\langle p^{-}\right\vert $ and $\hat{x}%
^{-}\,\left\vert p^{+}\right\rangle =-i\frac{\partial }{\partial p^{+}}%
\left\vert p^{+}\right\rangle $ consistent with the commutation rules. The
Fourier transform of these states correspond to diagonalizing the operators $%
\hat{x}^{+}$ and $\hat{x}^{-}$ in the position basis states $\left\langle
x^{+},p_{u},p_{v}\right\vert \equiv \left\langle x^{+}\right\vert $ and $%
\left\vert x^{-},p_{u},p_{v}\right\rangle \equiv \,\left\vert
x^{-}\right\rangle $, respectively. In the position basis $\hat{p}^{-}$ and $%
\hat{p}^{+}$ operate as $\left\langle x^{+}\right\vert \,\hat{p}^{-}=-i\frac{%
\partial }{\partial x^{+}}\left\langle x^{+}\right\vert $ and $\hat{p}%
^{+}\,\left\vert x^{-}\right\rangle =i\frac{\partial }{\partial x^{-}}%
\left\vert x^{-}\right\rangle $, respectively.

Now the matrix elements of the tachyon vertex operator can be evaluated in
the position or momentum basis just as one computes $D-$functions in a
representation of a group. To do that we find it convenient to compute the
vertex operator in the position basis as follows%
\begin{equation}
V_{x^{+},\,x^{-}}^{p_{v},\,p_{u},\,r}(g)=\left\langle x^{+}\right\vert
\,e^{a^{+}\hat{P}^{-}+a^{-}\hat{P}^{+}}e^{u\hat{J}+v\hat{T}}\,\left\vert
x^{-}\right\rangle \,,
\end{equation}%
since $\hat{P}^{-}$ has simple action on $\left\langle x^{+}\right\vert $,
and $\hat{P}^{+}$ has simple action on $\left\vert x^{-}\right\rangle $. We
use the same form also for the left/right vertex operators. In terms of
those the full vertex operator in the momentum basis is given by%
\begin{eqnarray}
V_{p^{-},\,p^{+}}^{p_{v},\,p_{u},\,r}(z,\bar{z}) &=&\left\langle
p^{-}\right\vert \,V\left( g_{L}\right) \,V\left( g_{R}\right) \,\left\vert
p^{+}\right\rangle  \notag \\
&=&\int \int d\tilde{p}^{+}d\tilde{p}^{-}\left\langle p^{-}\right\vert
\,V\left( g_{L}\right) \,\left\vert \tilde{p}^{+}\right\rangle \left\langle
\tilde{p}^{+}|\tilde{p}^{-}\right\rangle \left\langle \tilde{p}%
^{-}\right\vert V\left( g_{R}\right) \,\left\vert p^{+}\right\rangle  \notag
\\
&=&\int \int d\tilde{p}^{+}d\tilde{p}^{-}\left\langle \tilde{p}^{+}|\tilde{p}%
^{-}\right\rangle V_{p^{-},\,\tilde{p}^{+}}^{p_{v},\,p_{u},\,r}(z)\bar{V}_{%
\tilde{p}^{-},\,p^{+}}^{p_{v},\,p_{u},\,r}(\bar{z}).  \label{facVertex}
\end{eqnarray}%
Here again $\hat{P}^{-}$ has simple action on $\left\langle p^{-}\right\vert
$, and $\hat{P}^{+}$ has simple action on $\left\vert p^{+}\right\rangle $.

In the following the alternative representations of the tachyon vertex
operator in position and momentum bases will have different usages. The form
of the vertex operator in the position basis will be useful for comparison
to the wave functional (\ref{wfunction}) on the $H_{4}$ manifold, and for
possible interpretations of it in the BMN correspondence \cite{BMN}, as it
is done in \cite{BDM} for the case of AdS$_{3}$/CFT$_{2}$ correspondence.
Whereas the momentum basis form of the vertex operator will be useful in
determination of operator products of the vertex operator with the basic
fields in the theory. In fact we will show that with the correct quantum
ordering prescription the vertex operator has correct operator products with
the currents and it is a primary operator with conformal weight one. We will
also show that the conformal dimension comes in the correct form, being
proportional to the eigenvalue of the zero frequency Virasoro generator $%
L_{0}$. Due to its simpler form, the momentum basis vertex operator is also
expected to be useful in the computations of the correlation functions.

\subsection{Vertex Operator in the Position Basis}

We define the vertex operator in the position basis by%
\begin{equation}
V_{x^{+},\,x^{-}}^{p_{v},\,p_{u},\,r}(g)=\left\langle x^{+}\right\vert
\,e^{a^{+}\hat{P}^{-}+a^{-}\hat{P}^{+}}e^{u\hat{J}+v\hat{T}}\,\left\vert
x^{-}\right\rangle =\left\langle x^{+}\right\vert \,e^{a^{+}\hat{P}^{-}}e^{u%
\hat{J}+\left( v-ia^{+}a^{-}\right) \hat{T}}e^{a^{-}e^{iu}\hat{P}%
^{+}}\,\left\vert x^{-}\right\rangle .
\end{equation}%
Note that we are not using the factorized form of the vertex operator in
this basis. The vertex operator in the position basis is written in terms of
the coordinates of the group manifold. This is because this representation
of the vertex operator will turn out to be useful not in quantum
computations, but possible semi-classical interpretations like in \cite{BDM}.

In order to evaluate this matrix element of the vertex operator we note that
$\hat{P}^{-}\equiv i\hat{p}^{-}$ and $\hat{P}^{+}\equiv i\hat{p}^{+}$ behave
as translation operators, and $\hat{J}$ behaves as dilation operator,
together with the extra factor $e^{-iup_{v}}e^{-iu}$, in the position space.
Therefore we obtain%
\begin{eqnarray}
V_{x^{+},\,x^{-}}^{p_{v},\,p_{u},\,r}(g) &=&\left\langle
x^{+}+a^{+}\right\vert \,e^{u\hat{J}+\left( v-ia^{+}a^{-}\right) \hat{T}%
}\,\left\vert x^{-}-e^{iu}a^{-}\right\rangle \\
&=&e^{-i\left( up_{v}+\left( v-ia^{+}a^{-}\right) p_{u}\right)
}e^{-iu}\left\langle e^{-iu}\left( x^{+}+a^{+}\right)
|\,x^{-}-e^{iu}a^{-}\right\rangle \\
&=&e^{-i\left( up_{v}+\left( v-ia^{+}a^{-}\right) p_{u}\right)
}e^{-iu}\left\langle x^{+}+a^{+}|\,e^{-iu}\left( x^{-}-e^{iu}a^{-}\right)
\right\rangle .
\end{eqnarray}%
We must calculate the inner product $\left\langle x^{+}|\,x^{-}\right\rangle
$. Its properties under the dilations show that it is a function of only $%
x^{+}x^{-}$, therefore we write $\left\langle x^{+}|\,x^{-}\right\rangle
=f_{p_{u},\,r}(x^{+}x^{-})$. Then the full vertex operator is%
\begin{equation}
V_{x^{+},\,x^{-}}^{p_{v},\,p_{u},\,r}(g)=e^{-i\left( up_{v}+vp_{u}\right)
}e^{-a^{+}a^{-}p_{u}}\,e^{-iu}f_{p_{u},\,r}\left( e^{-iu}\left(
x^{+}+a^{+}\right) \left( x^{-}-e^{iu}a^{-}\right) \right) .
\end{equation}

In order to determine $f_{p_{u},\,r}(x^{+}x^{-})$ we first insert $\hat{P}%
^{-}$ and then $\hat{P}^{+}$\ in between left and right states and we find
their action on either states:%
\begin{eqnarray}
\left( -2p_{u}\,x^{-}+2\frac{r}{\partial _{-}}\right) f_{p_{u},\,r}
&=&\left\langle x^{+}\right\vert \,\hat{P}^{-}\,\left\vert
x^{-}\right\rangle =\partial _{+}\,f_{p_{u},\,r}  \label{xP-x} \\
-\partial _{-}\,f_{p_{u},\,r} &=&\left\langle x^{+}\right\vert \,\hat{P}%
^{+}\,\left\vert x^{-}\right\rangle =\left( 2p_{u}\,x^{+}-2\frac{r}{\partial
_{+}}\right) f_{p_{u},\,r}  \label{xP+x}
\end{eqnarray}%
Multiplying the equality (\ref{xP-x}) with $\partial _{-}$ and the equality (%
\ref{xP+x}) with $\partial _{+}$ we obtain differential equations%
\begin{eqnarray*}
\partial _{-}\partial _{+}\,f_{p_{u},\,r} &=&-2\left(
p_{u}+p_{u}\,x^{-}\partial _{-}-r\right) f_{p_{u},\,r} \\
\partial _{-}\partial _{+}\,f_{p_{u},\,r} &=&-2\left(
p_{u}+p_{u}\,x^{+}\partial _{+}-r\right) f_{p_{u},\,r}
\end{eqnarray*}%
Adding them and setting $x^{+}x^{-}=y$, we finally obtain the differential
equation%
\begin{equation}
y\partial _{y}^{2}f_{p_{u},\,r}\left( y\right) +\left( 1+2p_{u}y\right)
\partial _{y}f_{p_{u},\,r}\left( y\right) +\left( 2p_{u}-2r\right)
f_{p_{u},\,r}\left( y\right) =0.
\end{equation}%
This is the Kummer's differential equation. Its exact solution that is well
behaved at the origin is the confluent hypergeometric function%
\begin{equation}
f_{p_{u},\,r}\left( y\right) =e^{-2p_{u}y}\,_{1}F_{1}\left( 1+\frac{r}{p_{u}}%
,1;\,2p_{u}y\right) .
\end{equation}%
Then the vertex operator in the position basis becomes%
\begin{equation}
V_{x^{+},\,x^{-}}^{p_{v},\,p_{u},\,r}(g)=e^{-i\left( up_{v}+vp_{u}\right)
}e^{-p_{u}a^{+}a^{-}}e^{-iu}e^{-w}\,_{1}F_{1}\left( \alpha ,1;\,w\right) ,
\end{equation}%
where$\ \alpha =1+\frac{p^{+}p^{-}}{2p_{u}}$ and $w=2p_{u}e^{-iu}\left(
x^{+}+a^{+}\right) \left( x^{-}-e^{iu}a^{-}\right) $. This form of the
vertex operator resembles the wave functional (\ref{wfunction}). Therefore
for semi-classical considerations this form will have great utility. However
for quantum computations the confluent hypergeometric function would be
difficult to manipulate. For those kind of computations the momentum space
form of the vertex operator will be more useful, whose form is considerably
simpler as it will be seen in the next subsection.

\subsection{Vertex Operator in the Momentum Basis}

In this section we are going to describe only the holomorphic part of the
factorized vertex operator (\ref{facVertex}), $V_{p^{-},\,\tilde{p}%
^{+}}^{p_{v},\,p_{u},\,r}(z)$, in the momentum basis and show that it has
correct operator products with the holomorphic currents and the holomorphic
energy-momentum tensor. The anti-holomorphic part, $\bar{V}_{\tilde{p}%
^{-},\,p^{+}}^{p_{v},\,p_{u},\,r}(\bar{z})$, is insensitive to these
operator products and therefore we do not include it in the discussion. The
operator products in the anti-holomorphic sector has the same form as the
corresponding ones in the holomorphic sector. The operator products are
performed after determining the correct quantum ordering of the fields and
their zero modes in the expression for the vertex operator. We are going to
show that the holomorphic part of the vertex operator has the correct
conformal dimension, $h_{p_{v},\,p_{u},\,r}=p_{u}p_{v}+r+\frac{p_{u}}{2}.$

\subsubsection{Classical Expression}

The holomorphic part of the vertex operator in momentum basis is defined by%
\begin{equation}
V_{p^{+},\,p^{-}}^{p_{v},\,p_{u},\,r}(z)=\left\langle p^{-}\right\vert
\,e^{\theta ^{+}\hat{P}^{-}}e^{\phi \hat{J}+\varphi \hat{T}}e^{\theta ^{-}%
\hat{P}^{+}}\,\left\vert p^{+}\right\rangle .
\end{equation}%
In order to evaluate this we note that $\hat{P}^{-}\equiv i\hat{p}^{-}$ and $%
\hat{P}^{+}\equiv i\hat{p}^{+}$ are diagonal on states $\left\langle
p^{-}\right\vert $ and $\left\vert p^{+}\right\rangle $, respectively, and $%
\hat{J}$ behaves as dilation operator, together with the extra factor $%
e^{-i\phi p_{v}}$, in the momentum space. Therefore we obtain%
\begin{eqnarray}
V_{p^{+},\,p^{-}}^{p_{v},\,p_{u},\,r}(z) &=&e^{i\theta
^{+}p^{-}}\left\langle p^{-}\right\vert \,e^{\phi \hat{J}+\varphi \hat{T}%
}\,\left\vert p^{+}\right\rangle e^{i\theta ^{-}p^{+}}  \notag \\
&=&e^{i\theta ^{+}p^{-}}e^{-i\left( \phi p_{v}+\varphi p_{u}\right)
}\left\langle p^{-}|e^{i\phi }p^{+}\right\rangle e^{i\theta ^{-}p^{+}}
\notag \\
&=&e^{i\theta ^{+}p^{-}}e^{-i\left( \phi p_{v}+\varphi p_{u}\right)
}\left\langle e^{i\phi }p^{-}|p^{+}\right\rangle e^{i\theta ^{-}p^{+}}
\end{eqnarray}%
Now we define the function $\tilde{f}_{p_{u},\,r}\left( p^{+}p^{-}\right)
=\left\langle p^{-}|p^{+}\right\rangle $. This function must be a function
of the single variable $p^{+}p^{-}$ due to its properties under dilations.
Then the holomorphic part of the vertex operator is%
\begin{equation}
V_{p^{+},\,p^{-}}^{p_{v},\,p_{u},\,r}(z)=e^{i\theta ^{+}p^{-}}e^{-i\left(
\phi p_{v}+\varphi p_{u}\right) }\,\tilde{f}_{p_{u},\,r}\left( e^{i\phi
}p^{+}p^{-}\right) e^{i\theta ^{-}p^{+}}
\end{equation}

In order to determine $\tilde{f}_{p_{u},\,r}\left( p^{+}p^{-}\right) $ we
insert $\hat{P}^{-}$ and then $\hat{P}^{+}$\ in between left and right
states and then following the same procedure as in the previous subsection
we find the differential equation%
\begin{equation}
y\partial _{y}\tilde{f}_{p_{u},\,r}\left( y\right) +\left( \frac{r}{p_{u}}-%
\frac{y}{2p_{u}}\right) \tilde{f}_{p_{u},\,r}\left( y\right) =0.
\end{equation}%
where $y=p^{+}p^{-}$. The solution of this differential equation is found to
be $\tilde{f}_{p_{u},\,r}\left( y\right) =y^{-r/p_{u}}\exp \left(
y/2p_{u}\right) $. Then the holomorphic part of the vertex operator in the
momentum basis takes the form%
\begin{equation}
V_{p^{+},\,p^{-}}^{p_{v},\,p_{u},\,r}(z)=e^{i\theta ^{+}p^{-}}e^{-i\left(
\phi p_{v}+\varphi p_{u}\right) }\left( e^{i\phi }p^{+}p^{-}\right)
^{-r/p_{u}}\exp \left( e^{i\phi }\frac{p^{+}p^{-}}{2p_{u}}\right) e^{i\theta
^{-}p^{+}}.
\end{equation}

\subsubsection{Quantum Ordering and Operator Products}

To define the quantum expression of the vertex operator, we start by
preserving the order of the operators that comes from group theoretical
construction. Then we take the factors related to each non-commuting
generator of the algebra as already normal ordered:
\begin{eqnarray}
V_{p^{+},\,p^{-}}^{p_{v},\,p_{u},\,r}(z) &=&e^{i\theta
^{+}p^{-}}\left\langle p^{-}\right\vert \,\left( :e^{\phi \hat{J}+\varphi
\hat{T}}:\right) \,\left\vert p^{+}\right\rangle :e^{i\theta ^{-}p^{+}}: \\
&=&e^{i\theta ^{+}p^{-}}:e^{-i\left( \phi p_{v}+\varphi p_{u}\right) }\left(
e^{i\phi }p^{+}p^{-}\right) ^{-r/p_{u}}\exp \left( e^{i\phi }\frac{p^{+}p^{-}%
}{2p_{u}}\right) :  \notag \\
&&\times :\exp \left( ip^{+}\left( x_{0}^{-}+\frac{1}{2}\int^{z}dz^{^{\prime
}}\vartheta ^{-}(z^{^{\prime }})e^{i\phi (z^{^{\prime }})}\right) \right) :
\label{nonorderedV}
\end{eqnarray}%
where $\theta ^{-}\left( z\right) $ is written in terms of the canonical
variables (see equ.(\ref{thetaM})). Due to the definition of the canonical
variables, $e^{i\theta ^{+}p^{-}}$ does not need normal ordering. This gives
the ordering of the operators. However, in order to be able to use the Wick
theorem during the computations of the operator products we need to write
the quantum expression of the vertex operator in fully normal ordered form.
Using the contractions of free fields as given in equ. (\ref{fifi}-\ref{tete}%
) this fully normal ordered form is found as follows%
\begin{eqnarray}
V_{p^{+},\,p^{-}}^{p_{v},\,p_{u},\,r}(z) &=&:e^{i\theta
^{+}p^{-}}e^{-i\left( \phi p_{v}+\varphi p_{u}\right) }\left( e^{i\phi
}p^{+}p^{-}\right) ^{-r/p_{u}}\exp \left( e^{i\phi }\frac{p^{+}p^{-}}{2p_{u}}%
\right) e^{ip^{+}x_{0}^{-}}  \notag \\
&&\times \exp \left( i\frac{p^{+}}{2}\int^{z}dz^{^{\prime }}\left( \vartheta
^{-}(z^{^{\prime }})-i\frac{p^{-}}{z-z^{^{\prime }}}\right) e^{i\phi
(z^{^{\prime }})}(z-z^{^{\prime }})^{p_{u}}\right) :  \label{orderedV}
\end{eqnarray}%
Note that now the whole expression is between the normal ordering columns,
which is the reason of the complicated additional factors. Since the initial
definition (\ref{nonorderedV}) of the vertex operator is not fully normal
ordered, we should show that the vacuum expectation value of it is finite.
To compute this we use the last form (\ref{orderedV}) of the vertex operator
and find%
\begin{equation}
\left\langle 0\right\vert V_{p^{+},\,p^{-}}^{p_{v},\,p_{u},\,r}(z)\left\vert
0\right\rangle =\left( p^{+}p^{-}\right) ^{r/p_{u}}e^{p^{+}p^{-}/2p_{u}}\exp
\left( \frac{p^{+}p^{-}}{2}\int^{z}dz^{^{\prime }}(z-z^{^{\prime
}})^{p_{u}-1}\right)
\end{equation}%
which is finite.

The laborious technical details of the calculation of the operator products
of the currents and the energy-momentum tensor with the vertex operator is
similar to the computation presented in the appendix of \cite{BDM}.
Therefore, there is no need to repeat it here. We just would like to point
out a very important ingredient of the computation. This is the problem of
how the zero modes in the quantum expression for the vertex operator need to
be ordered.\ We found that the following ordering of the zero modes is the
correct one,%
\begin{equation}
\left[ \left\langle :e^{\phi \hat{J}+\varphi \hat{T}}:\right\rangle \right]
_{\text{zero modes}}=e^{-i\phi p_{v}}\left( e^{i\phi }p^{+}p^{-}\right)
^{-r/p_{u}}e^{\left( e^{i\phi }\frac{p^{+}p^{-}}{2p_{u}}\right)
}e^{-i\varphi p_{u}}\,
\end{equation}

With the ordering prescription given above, we found the correct operator
products with the holomorphic currents as%
\begin{equation}
\mathcal{J}_{L}^{i}\left( z\right) \times
V_{p^{+},\,p^{-}}^{p_{v},\,p_{u},\,r}(w)=\frac{1}{z-w}\left\langle
p^{-}\right\vert \,\mathcal{\hat{J}}^{i}V^{p_{v},\,p_{u},\,r}\,(w)\left\vert
p^{+}\right\rangle ,
\end{equation}%
where $\mathcal{\hat{J}}^{i}$ is $\hat{T},\ \hat{J},\ \hat{P}^{+}$ or $\hat{P%
}^{-}$. The action of $\mathcal{\hat{J}}^{i}$ on $\left\langle
p^{-}\right\vert $ is a differential operator that follows from the left
side of (\ref{operGen}). Thus, $\hat{P}^{-}=ip^{-}$, $\hat{J}=\,$dilations,
etc. We also found that the operator product of the vertex operator with the
energy-momentum tensor%
\begin{equation}
\mathcal{T}_{L}\left( z\right) \times
V_{p^{+},\,p^{-}}^{p_{v},\,p_{u},\,r}(w)=\frac{1}{(z-w)^{2}}\left\langle
p^{-}\right\vert \,\left( \frac{1}{2}L_{ij}\,\mathcal{\hat{J}}^{i}\,\mathcal{%
\hat{J}}^{j}\right) V^{p_{v},\,p_{u},\,r}\,(w)\left\vert p^{+}\right\rangle +%
\frac{1}{z-w}\partial _{w}V_{p^{+},\,p^{-}}^{p_{v},\,p_{u},\,r}(w)
\end{equation}%
gives the correct conformal dimension%
\begin{equation}
h_{p_{v},\,p_{u},\,r}=p_{u}p_{v}+r+\frac{p_{u}}{2}.
\end{equation}

This proves that the vertex operator we are proposing is the correct tachyon
vertex operator that correspond to the state $\left\vert 0,\vec{p}%
\right\rangle $ of the continuous series representation of $h_{4}$.

\section{Conclusions}

In this paper we have investigated the physical spectrum of the string
theory on NW spacetime. We started by showing that the NW spacetime is a
Penrose limit of AdS$_{2}\times $S$^{2}$ spacetime and its metric has the
form of an exact plane-wave background. NW spacetime also contains an
anti-symmetric field whose field strength is everywhere constant. The NW
spacetime can also be thought as the group manifold of the non-semi--simple
group $H_{4}$. This group is equivalent to either left or right part of the
isometry group. The corresponding algebra is the Heisenberg algebra with a
rotation operator added. From the isometry generators it is straightforward
to write the D'Alembertian on this background. We acted the D'Alembertian on
a wave functional written in a specific irreducible representation of $h_{4}$
and derived the dispersion relation.

The string theory on a group manifold is described by the WZNW\ model. We
realized the currents in the $H_{4}$ WZNW model in terms of four bosonic
free fields and constructed the string spectrum by using the negative
frequency modes of these free fields. The spectrum that is constructed by
the free field modes turned out to be equivalent to affine continuous series
representation of Kac--Moody algebra $\hat{h}_{4}$. We showed this by
checking that the eigenvalue of $L_{0}$ operator (half of the Hamiltonian)
on states is equivalent to the negative of the eigenvalue of the quadratic
Casimir operator in the continuous series representation of $h_{4}$. Due to
the indefinite signature of the metric there appeared the negative norm
states in the spectrum. However, it is argued that the situation is no worse
than the case in flat spacetime and the Virasoro constraints are enough to
eliminate all the negative norm states from the spectrum. We claimed that
the spectrum we found contains another spectrum as a sub-spectrum which is
obtained when one sets the transverse momenta to zero and thus gets the
affine discrete series representation as the spectrum of quantum strings. In
that case we had to explain why $p_{u}<1$ condition is not seen in our
construction. We argued that the mentioned condition is just an artifact of
the formalism used in the previous approaches \cite{KA},\cite{cheung}. Since
we realized the currents in terms of free bosons, there appeared logarithmic
cuts in the expressions of the currents. We required that in the physical
sector, the currents should be periodic due to the periodicity of $\sigma $
coordinate on closed string worldsheet. We found that this requirement
results in the quantization of the product of transverse momenta.

In the last section we revived the state-operator correspondence and
determined the tachyon vertex operator that corresponds to the $\left\vert 0,%
\vec{p}\right\rangle $ state in the string spectrum. We expressed the
classical form of the vertex operator in two different bases. The position
basis expression is given for the full vertex operator. This expression is
useful for comparison to the wave functional derived in the coherent state
basis in the continuous series representation of $h_{4}$. We expect that
this form of the vertex operator will also be useful to make claims about
the operators in the dual gauge theory as it is made in the case of AdS$_{3}$%
/CFT$_{2}$ correspondence in \cite{GKS},\cite{BDM}. The momentum basis
expression of the vertex operator is utilized for completely different
purpose. We observed that the momentum space form of the vertex operator
consists of simple exponentials and therefore for it a quantum ordering
prescription can be given more easily than the position space expression,
which contains confluent hypergeometric function. We determined the ordering
prescription and showed that the quantum ordered vertex operator has the
correct operator product expansions with the currents and the
energy--momentum tensor. From the latter OPE we obtained the conformal
dimension of the vertex operator in the expected form.

This way we have determined the physical spectrum of the string theory in
the NW\ spacetime. A possible extension of this work could be applying the
same ideas to higher dimensional $H_{2n+2}$ group manifolds. In the past
literature the irreducible representations of higher $h_{2n+2}$ algebras
were given in analogy with the $4$ dimensional case and therefore the
important continuous series representations of those higher algebras are
also missed \cite{bianchi}. We plan to investigate the construction of
string theory on $H_{2n+2}$ group manifolds again, with the knowledge of
these previously unknown representations. In view of the important changes
observed in the definition of the physical string spectrum in $H_{4}$ case,
we expect the same, if not more, new results to emerge in higher dimensional
cases.

The other channel of research, as mentioned in the introduction, is to
investigate the BMN dual gauge theory of the string theory we constructed.
Recalling the conjecture in \cite{stro} that the string theory on AdS$%
_{2}\times $S$^{2}$ is dual to a conformal quantum mechanics on the boundary
of AdS$_{2}$, we expect the bosonic string theory we quantize here would be
dual to some large N limit of conformal quantum mechanics on circle. We plan
to investigate this duality in the near future.

\section*{Acknowledgments}

C. D. is supported in part by the Turkish Academy of Sciences in the
framework of the Young Scientist Program (CD/T\"{U}BA--GEB\.{I}P/2002--1--7).


\begin{thebibliography}{99}
\bibitem{Nov82} S.~P.~Novikov,
Usp.\ Mat.\ Nauk \textbf{37N5} (1982) 3.

\bibitem{Wit84} E.~Witten, 
Commun.\ Math.\ Phys.\ \textbf{92} (1984) 455.

\bibitem{GW86} D.~Gepner and E.~Witten,
Nucl.\ Phys.\ B \textbf{278} (1986) 493.

\bibitem{ZF86} A.~B.~Zamolodchikov and V.~A.~Fateev,
Sov.\ J.\ Nucl.\ Phys.\ \textbf{43} (1986) 657 [Yad.\ Fiz.\ \textbf{43}
(1986) 1031].

\bibitem{BRFW} J. Balog, L. O'Raighfertaigh, P. Forgacs, and A. Wipf, Nucl.
Phys. \textbf{B325}, 225 (1989).

\bibitem{B95} I. Bars, Phys. Rev. \textbf{D53}, 3308 (1996)
[arXiv:hep-th/9503205]. \newline
I. Bars, \textit{``Solution of the SL(2,R) String in Curved Spacetime''}, in
Proceedings of Strings'95 Conference, \textit{Future Perspectives in String
Theory}, Eds. I. Bars et al., World Scientific (1996), page 3
[arXiv:hep-th/9511187].

\bibitem{BDM} I. Bars, C. Deliduman and D. Minic, \textit{``String Theory on
AdS}$_{3}$\textit{\ Revisited''}, [arXiv:hep-th/9907087].

\bibitem{NW} C.~R.~Nappi and E.~Witten,
Phys.\ Rev.\ Lett.\ \textbf{71}, 3751 (1993) [arXiv:hep-th/9310112].

\bibitem{BMN} D.~Berenstein, J.~M.~Maldacena and H.~Nastase,
JHEP \textbf{0204} (2002) 013 [arXiv:hep-th/0202021].

\bibitem{penrose} R. Penrose, \textit{``Any Spacetime Has a Plane Wave as a
Limit,''} in \textit{Differential Geometry and Relativity}, Reidel,
Dordrecht, 1976, pp. 271-275.

\bibitem{guven} R.~Gueven, 
Phys.\ Lett.\ B \textbf{482} (2000) 255 [arXiv:hep-th/0005061].

\bibitem{blau} M.~Blau, J.~Figueroa-O'Farrill and G.~Papadopoulos,
Class.\ Quant.\ Grav.\ \textbf{19} (2002) 4753 [arXiv:hep-th/0202111].

\bibitem{stro} A.~Strominger, 
JHEP \textbf{0403} (2004) 066 [arXiv:hep-th/0312194].

\bibitem{KK} E.~Kiritsis and C.~Kounnas,
Phys.\ Lett.\ B \textbf{320}, 264 (1994) [Addendum-ibid.\ B \textbf{325},
536 (1994)] [arXiv:hep-th/9310202].

\bibitem{KKL} E.~Kiritsis, C.~Kounnas and D.~Lust,
Phys.\ Lett.\ B \textbf{331}, 321 (1994) [arXiv:hep-th/9404114].

\bibitem{forgacs} P.~Forgacs, P.~A.~Horvathy, Z.~Horvath and L.~Palla,
Heavy Ion Phys.\ \textbf{1} (1995) 65 [arXiv:hep-th/9503222].

\bibitem{KA} G.~D'Appollonio and E.~Kiritsis,
Nucl.\ Phys.\ B \textbf{674} (2003) 80 [arXiv:hep-th/0305081].

\bibitem{cheung} Y.~K.~Cheung, L.~Freidel and K.~Savvidy,
JHEP \textbf{0402}, 054 (2004) [arXiv:hep-th/0309005].

\bibitem{bianchi} M.~Bianchi, G.~D'Appollonio, E.~Kiritsis and O.~Zapata,
JHEP \textbf{0404}, 074 (2004) [arXiv:hep-th/0402004].

\bibitem{KP} E.~Kiritsis and B.~Pioline,
JHEP \textbf{0208} (2002) 048 [arXiv:hep-th/0204004].

\bibitem{macars} P.~Forgacs, Z.~Horvath and L.~Palla, unpublished.

\bibitem{HSPRD} G.~T.~Horowitz and A.~R.~Steif,
Phys.\ Rev.\ D \textbf{42} (1990) 1950.

\bibitem{HS} A.~Hatzinikitas and I.~Smyrnakis, ``Closed bosonic string
partition function in time independent exact pp-wave background,''
arXiv:hep-th/0303043.

\bibitem{brinkmann} H. W. Brinkmann, Math. Ann. \textbf{18} (1925) 119.

\bibitem{KSMH} H.~Stephani, D.~Kramer, M.~MacCallum, C.~Hoenselaers and
E.~Herlt, \textit{``Exact solutions of Einstein's field equations,''}
Cambridge Univ. Press (2003).

\bibitem{baldwin} O. R. Baldwin and G. B. Jeffrey, Proc. Roy. Soc. Lond. A
\textbf{111} (1926) 95.

\bibitem{KPR} C.~Kounnas, M.~Porrati and B.~Rostand,
Phys.\ Lett.\ B \textbf{258} (1991) 61. \newline
C.~G.~Callan, J.~A.~Harvey and A.~Strominger,
Nucl.\ Phys.\ B \textbf{367} (1991) 60.

\bibitem{GO} J.~Gomis and H.~Ooguri,
Nucl.\ Phys.\ B \textbf{635} (2002) 106 [arXiv:hep-th/0202157].

\bibitem{S98} A.~Strominger, 
JHEP \textbf{9901} (1999) 007 [arXiv:hep-th/9809027].

\bibitem{KlPe} J. R. Klauder, J. Math. Phys. \textbf{4} (1963) 1058. \newline
A.~M.~Perelomov, \textit{``Generalized Coherent States And Their
Applications,''} Springer (1986).

\bibitem{BR} A. O. Barut and L. Girardello, Commun. Math. Phys. \textbf{21}
(1971) 41.

\bibitem{HK} M.~B.~Halpern and E.~Kiritsis,
Mod.\ Phys.\ Lett.\ A \textbf{4} (1989) 1373.

\bibitem{Moh93} N.~Mohammedi,
Phys.\ Lett.\ B \textbf{325} (1994) 371 [arXiv:hep-th/9312182].

\bibitem{thorn} C.~B.~Thorn,
Nucl.\ Phys.\ B \textbf{248} (1984) 551.

\bibitem{GKS} A.~Giveon, D.~Kutasov and N.~Seiberg,
Adv.\ Theor.\ Math.\ Phys.\ \textbf{2} (1998) 733 [arXiv:hep-th/9806194].
\end{thebibliography}
\end{document}